\documentclass[submission, Phys]{SciPost}
\usepackage{color,amsmath,amssymb,multirow,hyperref,booktabs,graphicx,mathtools,xspace,hhline,longtable,makecell} 
\newcommand{\arxiv}[1]{\href{http://arxiv.org/abs/#1}{arXiv:#1}}
\newcommand{\urlx}[1]{\href{#1}{#1}}

\newcommand\one{\leavevmode\hbox{\small1\normalsize\kern-.33em1}}

\newcommand{\lag}{\mathcal{L}}

\newcommand{\qqquad}{\qquad \qquad}

\newcommand{\met}{\slashchar{E}_T}

\newcommand{\gev}{\text{GeV}}

\def\slashchar#1{\setbox0=\hbox{$#1$}           
   \dimen0=\wd0                                 
   \setbox1=\hbox{/} \dimen1=\wd1               
   \ifdim\dimen0>\dimen1                        
      \rlap{\hbox to \dimen0{\hfil/\hfil}}      
      #1                                        
   \else                                        
      \rlap{\hbox to \dimen1{\hfil$#1$\hfil}}   
      /                                         
   \fi}

\newcommand{\eg}{\textsl{e.g.}\;}

\newcommand{\be}{\begin{eqnarray*}}
\newcommand{\ee}{\end{eqnarray*}}

\newcommand{\bee}{\begin{eqnarray}}
\newcommand{\eee}{\end{eqnarray}}
\newcommand{\beeq}{\begin{equation}}
\newcommand{\eeeq}{\end{equation}}

\begin{document}

\begin{center}{\Large \textbf{
Quark-Gluon Tagging: Machine Learning vs Detector
}}\end{center}

\begin{center}
Gregor Kasieczka\textsuperscript{2},
Nicholas Kiefer\textsuperscript{1},
Tilman Plehn\textsuperscript{1}, and 
Jennifer M. Thompson\textsuperscript{1}
\end{center}

\begin{center}
{\bf 1} Institut f\"ur Theoretische Physik, Universit\"at Heidelberg, Germany\\
{\bf 2} Institut f\"ur Experimentalphysik, Universit\"at Hamburg, Germany
plehn@uni-heidelberg.de
\end{center}

\begin{center}
\today
\end{center}


\section*{Abstract}
{\bf Distinguishing quarks from gluons based on low-level detector
  output is one of the most challenging applications of multi-variate
  and machine learning techniques at the LHC. We first show the
  performance of our 4-vector-based LoLa tagger without and after
  considering detector effects. We then discuss two benchmark
  applications, mono-jet searches with a gluon-rich signal and di-jet
  resonances with a quark-rich signal. In both cases an immediate
  benefit compared to the standard event-level analysis exists.}

\vspace{10pt}
\noindent\rule{\textwidth}{1pt}
\tableofcontents\thispagestyle{fancy}
\noindent\rule{\textwidth}{1pt}
\vspace{10pt}

\newpage
\section{Introduction}
\label{sec:intro}

Since the start of the LHC our view of jets as analysis objects has
fundamentally changed. While jets with reconstructed 4-momenta
matching hard partons still serve as the key objects of essentially
all analyses, their internal structure can now be exploited
systematically. In that sense, jets merely define the boundary between
event-level observables and subjet observables. The subjet aspect is
currently undergoing a paradigm change: rather than defining
high-level kinematic observables for the jet constituents and
analyzing them using multivariate methods, we can use modern machine
learning approaches to analyze low-level detector outputs like the
measured 4-vectors entries directly~\cite{ml_review}. For this
low-level input we employ modern machine learning techniques, usually
advertized with the term \textsl{deep learning}.

Theoretically and experimentally well-controlled applications of
machine learning in subjet physics include hadronic
$W/Z$-jets~\cite{images,irvine_w,aussies,gan,kyle,information}, Higgs
jets~\cite{higgsbb,mihoko}, top
jets~\cite{ann_top,deeptop,canadians,lola,top_lstm,shih,karl,top_ir},
or model-independent searches for hard new physics in
jets~\cite{autoencoder} quark--gluon discrimination has a long
history~\cite{qg_early,qg_ml,cleo,opal,delphi} and is used at the
LHC~\cite{qg_atlas,qg_cms}. However, distinguishing quark and gluon
jets poses serious theoretical and simulational challenges, like, that
they are not defined in QCD beyond tree
level~\cite{qg_theo,qg_review,systematics,qg_operational,deep_sets}. Nevertheless, efficient
machine learning approaches have been devised to separate `quark jets'
from `gluon
jets'~\cite{in_color,qg_rnn,qg_fnn,impure,charge,heavyion,cwola,qg_image,tth}.
One way we can overcome the fundamental problems in defining quark and
gluon jets is to instead ask for a well-defined hypotheses in terms of
LHC signatures, involving mostly gluons vs gluons in the signal and
background processes~\cite{pure_samples,qg_ben,onke,sakaki}.

Before we employ modern machine learning to separate processes with
mostly hard quarks from those with mostly hard gluons we review the
known high-level variables. Unlike for many other subjet analyses
these observables rely on tracking information with its excellent
resolution, and cannot be considered infrared-safe observables or
easily interpretable in perturbative
QCD~\cite{qg_theo,qg_review,systematics}. When we switch to low-level
inputs this means that we cannot hope for the calorimeter resolution
to provide a generous binning and to render us insensitive to
additional detector effects. Moreover, any promising network
architecture needs to combine standard calorimeter images and tracking
information with its vastly better angular
resolution~\cite{in_color}. We will use our 4-vector-based
\textsc{LoLa} framework developed for top tagging including
calorimeter and tracking information~\cite{lola} to extract the
necessary information from measured particle-flow objects and to
quantify the sensitivity to soft tracks in the detector.  The latter
is especially relevant when we benchmark the machine learning approach
compared to a multi-variate analysis of the traditional quark--gluon
variables. In Sec.~\ref{sec:qg} we analyze idealistic, pure quark and
gluon samples to benchmark our tagger in the presence of detector
effects~\cite{qg_rnn}, to compare its performance to the classic
quark--gluon variables, and to study the correlation with the jet
momentum.

Finally, we will establish realistic and relevant benchmark analysis
for quark--gluon tagging at the LHC. Unfortunately, it is already
known that quark--gluon tagging does not significantly improve
weak-boson-fusion analyses at the LHC~\cite{onke}. Two often-discussed
candidate analyses for quark-gluon tagging in LHC searches are
\begin{enumerate}
\setlength\itemsep{0em}
\item mono-jet dark matter searches with a gluon-dominated signal,
  Sec.~\ref{sec:mono}, and
\item di-jet resonance searches with a
  quark-dominated signal, Sec.~\ref{sec:resonance}.
\end{enumerate}
For both cases we motivate the use of quark--gluon tagging, show how
our \textsc{LoLa} tagger helps extract the signal, and discuss the
limitations in a realistic analysis setup.

\section{Ideal world} 
\label{sec:qg}

In spite of the fact that a parton-level definition of quark and
gluons becomes ambiguous beyond leading-order QCD, we start with an
analysis of jets coming from hard quarks and gluons at tree-level and
based on Monte Carlo truth. The impact of this simplification should
eventually be tested including higher-order effects. At this point it
will allow us to identify the leading subjet properties of such jets
and to compare our deep learning approach with established approaches.

We generate quark and gluon jet samples using di-jet events with
\textsc{Sherpa}2.2.1~\cite{sherpa} at 14~TeV.  We do not simulate any
multiple interactions and any effects from pile up could be dealt with
by using established techniques as well as recently proposed
tools~\cite{pile_up}. For quark jets we extract the subprocesses
$gg/q\bar q \to q\bar q$ and $qq \to qq$, for the gluon jets we keep
the subprocesses $gg/q\bar q \to gg$. We pass these events through
\textsc{Delphes}3.3.2~\cite{delphes}, using the standard ATLAS
card. Finally, we cluster the particle flow
objects~\cite{particleflow} into anti-$k_T$~\cite{anti_kt} jets of
radius $R = 0.4$ using \textsc{FastJet}3.1.3~\cite{fastjet}. All jet
constituents have to be central in the detector, with $|\eta| < 2.5$
and $p_T > 1$~GeV.  Unless explicitly mentioned, our jets have
\begin{align}
p_{T,j} = 200~...~220~\gev \; . 
\end{align}
This setup closely follows Ref.~\cite{in_color}, with an additional
fast detector simulation. We do, however, find that switching from
\textsc{Pythia} to \textsc{Sherpa} makes quark--gluon discrimination
generally a little harder~\cite{systematics}.

\subsection{Standard observables}
\label{sec:obs}

Distinguishing quark jets from gluon jets exploits two
features~\cite{casimir_poisson}: first, radiating a gluon off a hard
gluon versus off a hard quark comes with a ratio of color factors
$C_A/C_F = 9/4$. This leads to a higher particle multiplicity
($n_\text{PF}$) and a broader radiation distribution or girth
($w_\text{PF}$)~\cite{qg_dist:wPF} for hard gluons; second, the
splitting functions $\hat P_{gg}(z)$ and $\hat P_{qq}(z)$ differ in
the soft limits. The harder fragmentation for quarks makes quark jet
constituents carry a larger average fraction of the jet energy,
tracked by the variable $p_T D$~\cite{qg_cms}.  In addition, the
two-point energy correlator $C_{0.2}$ separates quarks and gluons with
an optimized power of $\Delta R_{ij}$~\cite{qg_dist:C}.  This allows
us to define the four established observables
\begin{align}
n_\text{PF} &=  \sum_i 1 \qqquad &
w_\text{PF} &= \frac{\sum_i p_{T,i} \Delta R_{i,\text{jet}}}{ \sum_i p_{T,i}}  
\notag \\
p_T D &= \frac{\sqrt{\sum_i p_{T,i}^2 }}{ \sum_i p_{T,i} } \qqquad &
C_{0.2} &= \frac{\sum_{ij} E_{T,i} E_{T,j} (\Delta R_{ij})^{0.2} }{\sum_i E_{T,i}^2} \; .
\label{eq:qg_obs}
\end{align}
In addition, we evaluate
the highest fraction of $p_{T,\text{jet}}$ contained in a single jet
constituent~\cite{qg_dist:xmax}, and the minimum number of
constituents which contain 95\% of
$p_{T,\text{jet}}$~\cite{qg_dist:n90},
\begin{align}
x_\text{max} 
\qqquad \text{and} \qqquad 
N_{95} \; .
\end{align}
The latter is obviously correlated with the number of constituents
$n_\text{PF}$. All jet constituents summed over are defined as Delphes
E-flow objects, combining both the calorimeter and the tracking
information.

\begin{figure}[t]
\includegraphics[width=0.328\textwidth]{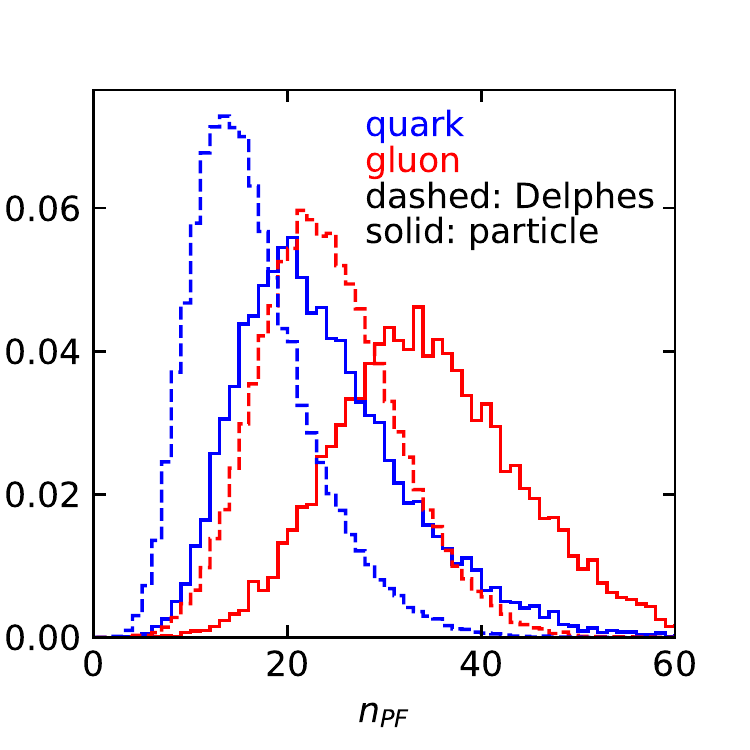}
\includegraphics[width=0.328\textwidth]{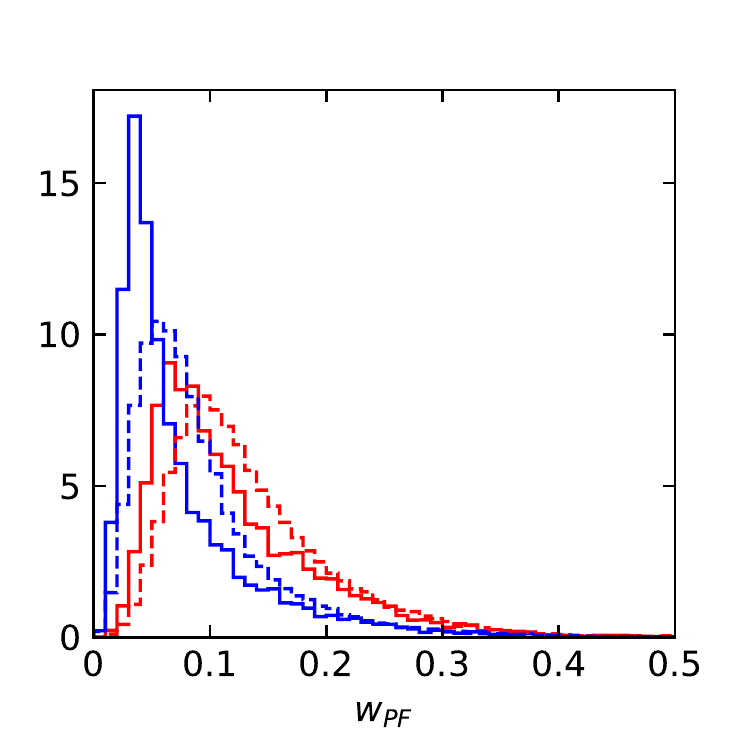}
\includegraphics[width=0.328\textwidth]{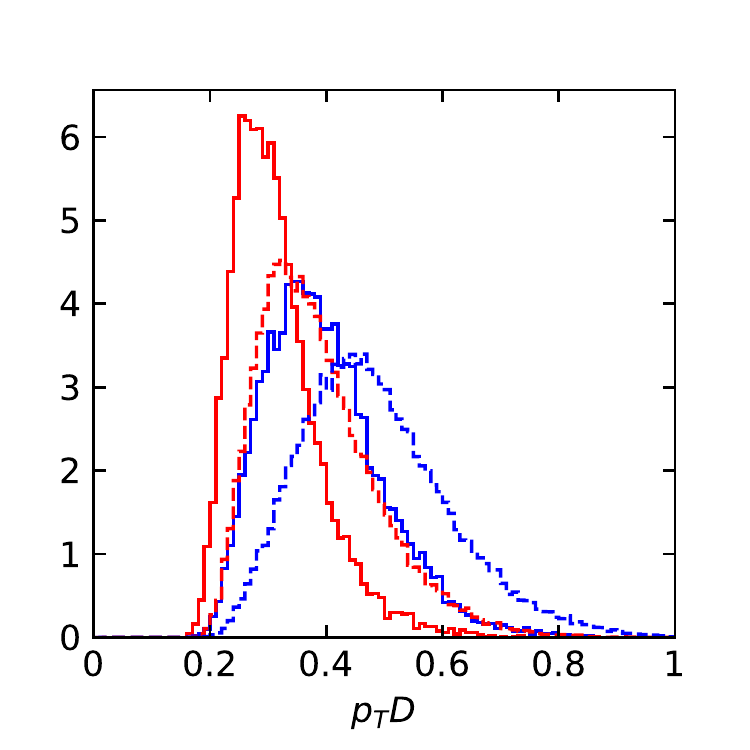} \\
\includegraphics[width=0.328\textwidth]{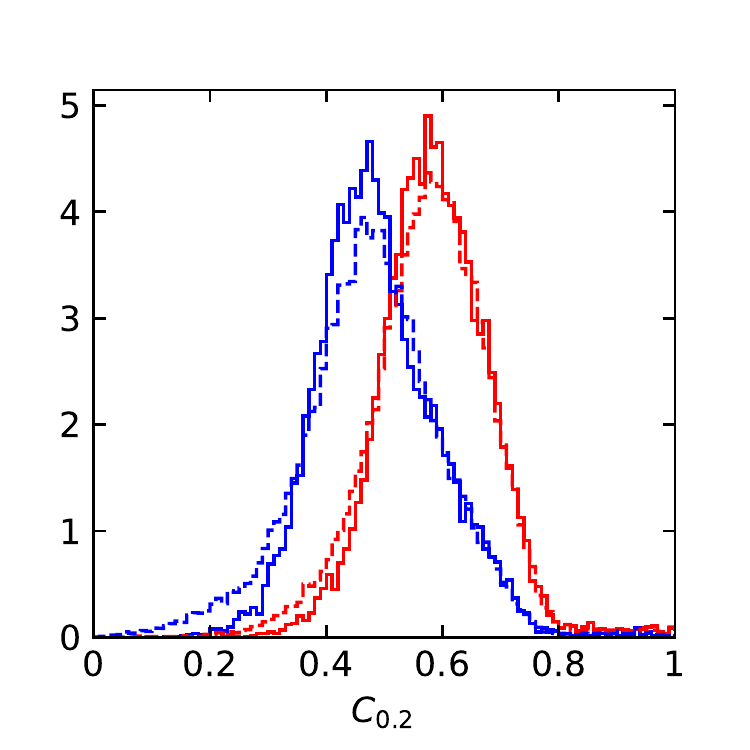}
\includegraphics[width=0.328\textwidth]{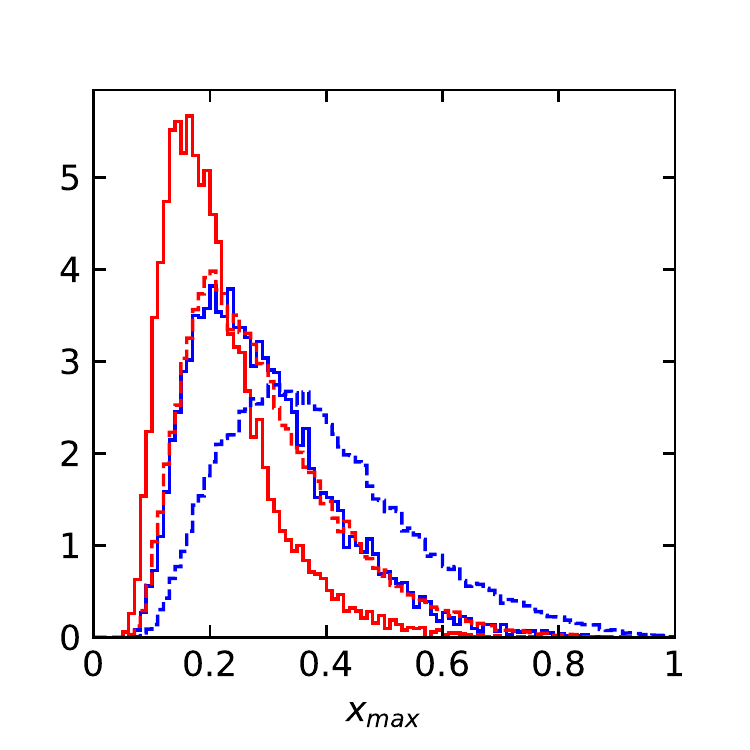}
\includegraphics[width=0.328\textwidth]{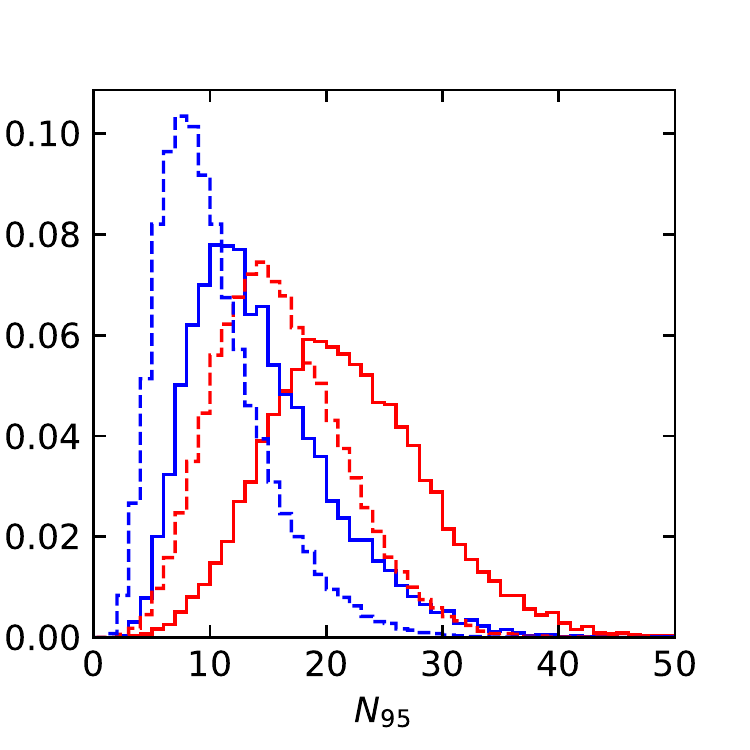}
\caption{Normalized distributions for the subjet variables described in the text
  for pure quark and pure gluon jets, with and without detector
  effects. Jets are selected with $p_T =
  200~...~220$~GeV.}
\label{fig:subjet_vars}
\end{figure}

Distributions of all these
observables for pure quark and gluon samples are shown in
Fig.~\ref{fig:subjet_vars}, both in an ideal setup and at the level of
particle flow object after fast detector simulation. The IR-sensitive
and theoretically challenging observable $n_\text{PF}$ shows large
differences because LHC detectors rapidly lose sensitivity for soft
constituents.  The $p_TD$ distribution is similarly sensitive.  When we
add a soft constituent we find that the numerator and denominator
change differently,
\begin{align}
p_T D 
\sim \frac{\sqrt{p_T^2 + \epsilon^2 p_T^2}}{p_T + \epsilon p_T}
\approx \frac{1 + \epsilon^2/2}{1+ \epsilon} \; .
\end{align}
This way $p_T D$ shifts towards smaller values, which do not survive
a detector simulations, as seen in Fig.~\ref{fig:subjet_vars}. The
situation is more stable for the $p_T$-weighted $w_\text{PF}$ and for
$C_{0.2}$.

\begin{figure}[t]
\includegraphics[width=0.47\textwidth]{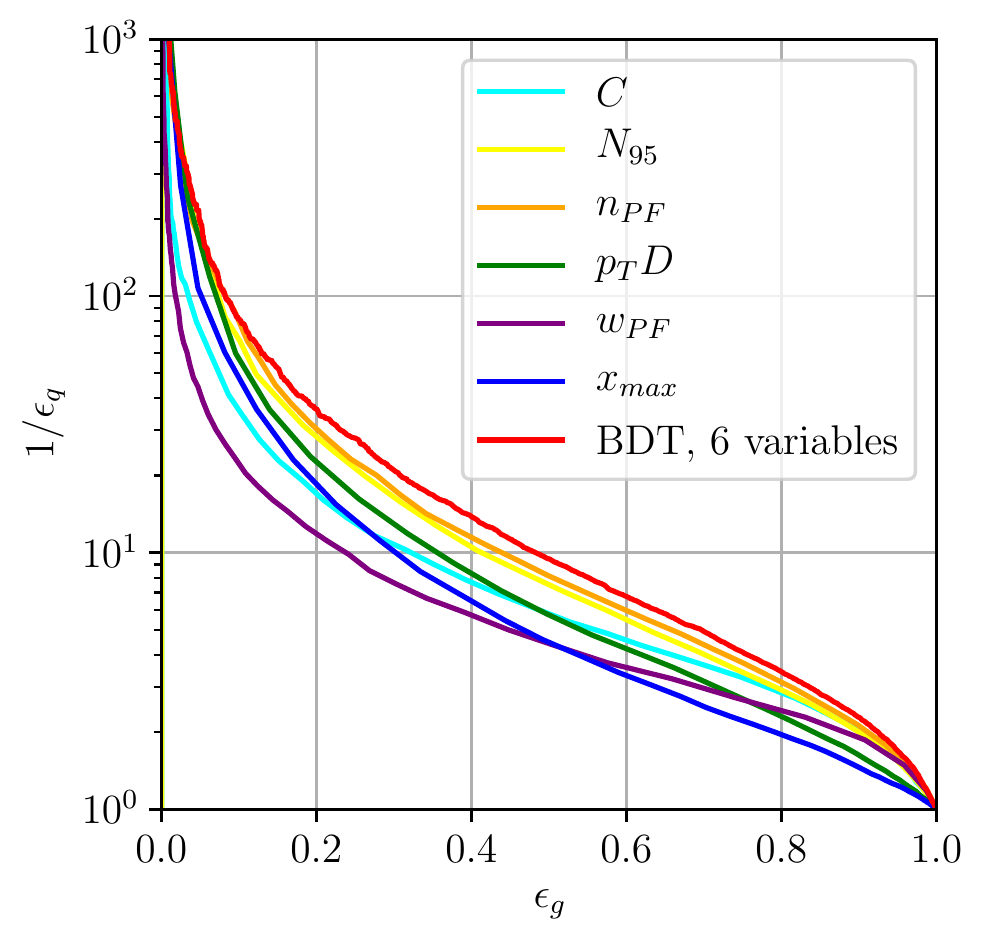}
\hspace*{0.04\textwidth}
\includegraphics[width=0.47\textwidth]{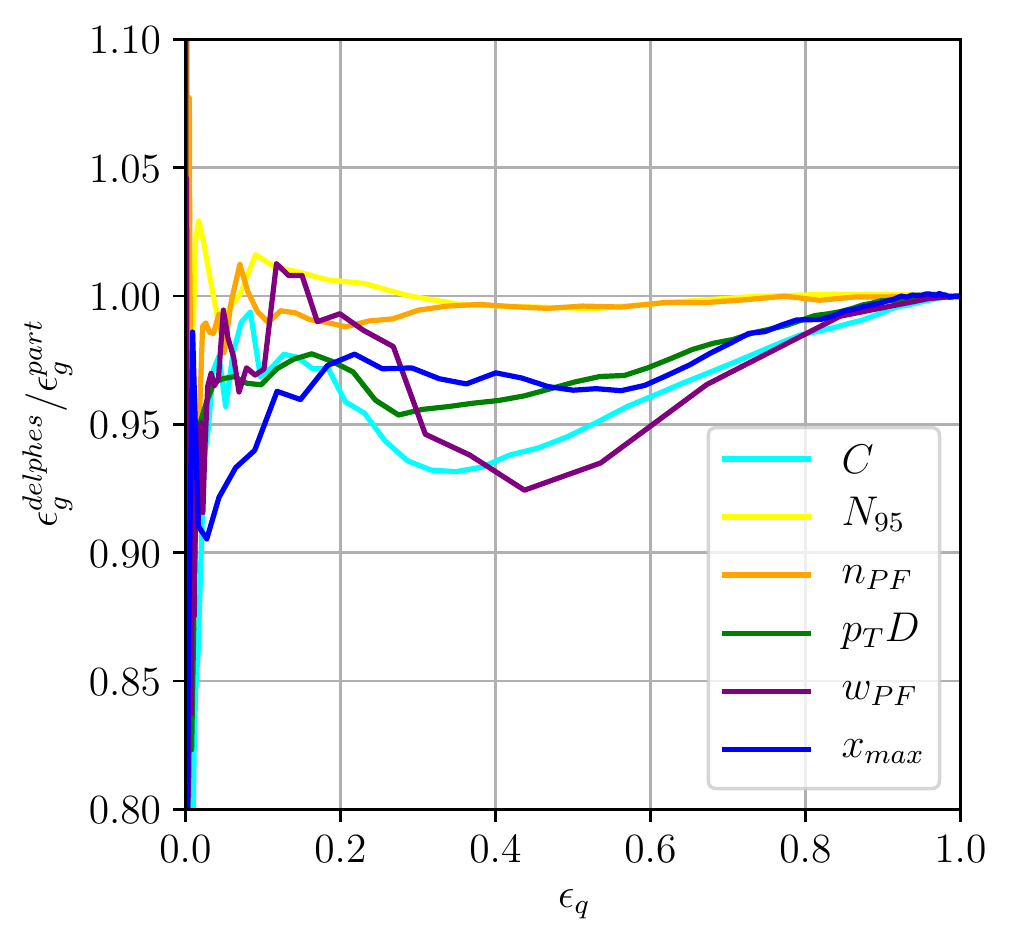}
\caption{Left: ROC curves for the six quark--gluon observables
  discussed in the text, including a combination through a BDT,
  without detector effects. Right: detector effect illustrated as
  ratios of single-observable ROC curves, shown as the ratio
  $\epsilon^\textsc{Delphes}_g/\epsilon^\text{particles}_g$.}
\label{fig:bdt_rocs}
\end{figure}

The individual performance of these six observables in tagging pure
quark and gluon jets without detector effects is illustrated in the
left panel of Fig.~\ref{fig:bdt_rocs}. Each of the observables indeed
contributes to quark-gluon discrimination. The number of constituents
$n_\text{PF}$ is the most powerful single variable, with almost
identical performance to $N_{95}$. This confirms the findings of
Ref.~\cite{in_color} in the absence of detector effects. To maximize
their separation power we combine all six of them into a boosted
decision tree (BDT), implemented in \textsc{Scikit-Learn} using a
gradient boosting classifier with 50 estimators, a maximum tree depth
of 4, a sub-sampling fraction of 0.9 and a learning rate of 1. The
classifier is trained on a sample of 500k quark and gluon jets, 5\% of
which are set aside as a test sample. The corresponding ROC curves are
also shown in Fig.~\ref{fig:bdt_rocs}, showing a small improvement
over the most powerful, but poorly defined variable $n_\text{PF}$.  In
the right panel of Fig.~\ref{fig:bdt_rocs} we compare the ROC curves
with and without detector simulation. From Fig.~\ref{fig:subjet_vars}
we know that for all variables the detector affects the quark and
gluon distributions systematically, both shifting and broadening the
features.  We can quantify the detector effect for instance by
comparing the gluon tagging efficiencies with and without
\textsc{Delphes} as a function of the quark efficiency in the right
panel of Fig.~\ref{fig:bdt_rocs}. The result suffers from numerical
fluctuations for extremely small $\epsilon_q < 0.01$, but for the bulk
of the ROC curves for each observable the detector effect are within
10\% of the ideal curve. Interestingly, the simplest observables
$n_\text{PF}$ and $N_{95}$ turn out the most stable in distinguishing
quarks from gluons. This suggest that they offer sizeable quark-gluon
separation power already in phase space regions which are not affected
by detector effects.

\begin{figure}[t]
\centering
\includegraphics[width=0.47\textwidth]{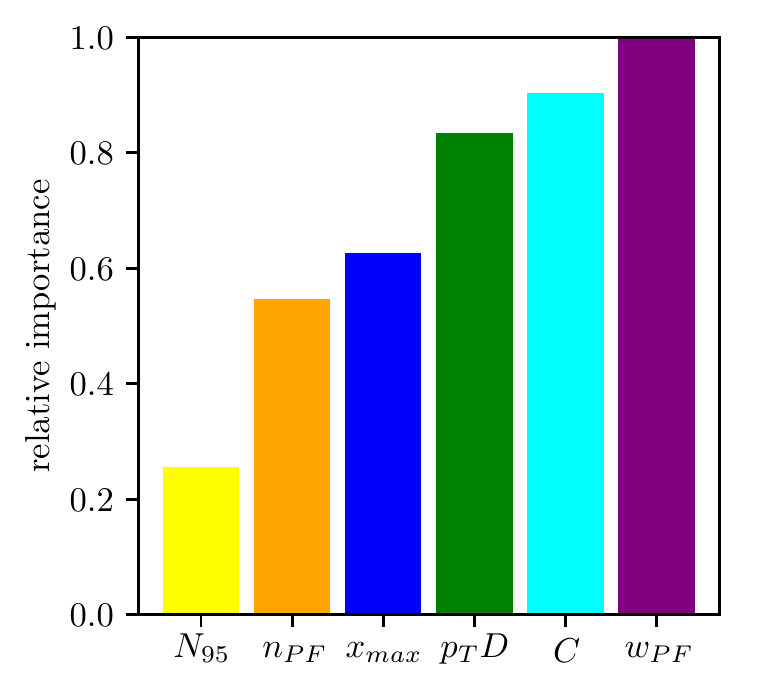}
\caption{Feature importance of each variable in the BDT, after a
  \textsc{Delphes} simulation, normalized to the most important
  feature.}
\label{fig:bdt_imp}
\end{figure}

Given that our six jet observables are an ad-hoc collection and do not
form any kind of basis in a space of correlators, it is neither
guaranteed that they include all available information nor that they
form a minimal set. The first question can be answered when we
eventually compare their separation power to our deep-learning
tagger. To tackle the second question we plot the feature importance
of each input variable in Fig.~\ref{fig:bdt_imp}. For a
variable $x$ we want to look at individual nodes $t$ making up a tree
and how often $x$ is used for a split $s_t$.  For each split we first
compute the probability $p(t)$ for a sample to reach the node $t$ and
define the purity of each node by the Gini index
\begin{align}
i(t) = 1- \sum_\text{outcomes $j$} p_j^2(t) 
     = 2 \, p_1(t) p_2(t) < \frac{1}{2} \; ,
\end{align}
where the last step holds for two classification hypotheses and gives
twice the probability of choosing a data point of category $j$ at node
$t$, multiplied by the probability of mis-labeling it. It reaches its
maximum for even probabilities and tends to zero if all the samples in
a node are of the same category. In that sense it is a measure of the
purity or impurity of the sample at node $t$. 
Next, we compute the change in purity of the node $t$ when we define a
split $s_t$ in terms of the variable $x$, defining $\Delta i(s_t, t)$.
This allows us to quantify the importance of a variable $x$ as
\begin{align}
\text{Imp}(x) \propto \sum_\text{trees} \sum_\text{nodes} p(t) \, \Delta i(t,x) \; ,
\end{align}
modulo a normalization constant.  A decision tree is essentially a
series of nodes which splits the samples such that the decrease in
impurity is maximized, hence more important features are more often
used to split the samples. Because cutting on a one-dimensional
distribution as shown in Fig.~\ref{fig:bdt_rocs} masks correlations,
the importance allows us to define the feature that best separates the
data whilst being least correlated with other variables.  We show the
results in Fig.~\ref{fig:bdt_imp} and find a start constrast to the
single-variable results of Fig.~\ref{fig:bdt_rocs}. The most powerful
single observables $n_\text{PF}$ and $N_{95}$ are strongly correlated
with the leading variable $w_\text{PF}$ and therefore contribute
little to the multi-variate analysis. Instead, the two-point
correlation $C_{0.2}$, which carries extra information than the other
(first-order moment) variables, is the most important additional
feature. Amusingly, these two leading observables $w_\text{PF}$ and
$C_{0.2}$ are also IR-safe~\cite{qg_dist:C}.  All other observables
constribute to the quark-gluon separation, but with different impact.

We close with a word of caution. The subjet observables given in
Eq.\eqref{eq:qg_obs} are not theoretically well-defined observables
which we can compute based on QCD. Instead, they are statistical
descriptions of jet constituents, including two-object correlators, in
some cases IR-modified by an appropriate energy scaling.  Relying on
not consistently IR-safe observables complicates quark-gluon
separation at the LHC, but does not make it
impossible~\cite{qg_theo,qg_review,systematics,deep_sets}.  The main problem is
that we cannot define quark or gluon jets in perturbative QCD or in
Monte-Carlo simulations beyond leading order in QCD. Clearly, these
observables as well as low-level observables cannot be directly used
to study QCD properties of subjets. On the other hand.  IR-safety does
not have to be an issue for data-to-data analyses, like quark-gluon
tagging trained on observed jets. All we need to do is define the
quark and gluon labels in relation to a hard process which predicts
mostly quarks or mostly gluons, rather than jet by
jet~\cite{qg_operational}. This way we can use the potentially
powerful soft and collinear subjet information as long as we do not
attempt to interpret these measurements in terms of QCD.

\subsection{Charging LoLa}
\label{sec:pure}

Given our result for the multi-variate analysis of high-level
substructure variables, it is natural to ask what happens when we
attempt to capture all available information from low-level
observables using a deep neural network. To combine information from
the calorimeter and the tracker with its different resolution, a
promising approach is the \textsc{LoLa} architecture applied to
particle flow objects, developed for the \textsc{DeepTopLoLa}
tagger~\cite{lola}. The input to the network are the $N$ jet
constituent 4-vectors sorted by $p_T$, 
\begin{align}
( k_{\mu,i} ) = 
\begin{pmatrix}
k_{0,1} &  k_{0,2} & \cdots &  k_{0,N} \\
k_{1,1} &  k_{1,2} & \cdots &  k_{1,N} \\ 
k_{2,1} &  k_{2,2} & \cdots &  k_{2,N} \\ 
k_{3,1} &  k_{3,2} & \cdots &  k_{3,N}  
\end{pmatrix} \; .
\label{eq:def_input}
\end{align}
Since $N$ varies from jet to jet, we zero-pad jets with fewer than
$N$ constituents, and increase $N$ until the tagging performance is
saturated, for most physics scenarios giving $N = 25~...~30$. Above
this the soft jet constituents carry too little information to
compensate for the increasing computation time. Inspired by the
structure of recombination jet algorithms, we multiply the original
4-vectors with a trainable matrix $C_{ij}$, defining a
combination layer (CoLa)
\begin{align}
k_{\mu,i} \stackrel{\text{CoLa}}{\longrightarrow}
\widetilde{k}_{\mu,j} 
&= k_{\mu,i} \; C_{ij} \notag \\
\quad \text{with} \quad
C &= 
 \begin{pmatrix}
1 				&  1  &  \cdots  &  0      	&  \chi_{1}  &  \cdots  &  0  	& C_{1,N+2} & \cdots & C_{1,M} \\[-2mm]
\vdots 		&  & \ddots &				&  &  \ddots  &						& \vdots & \ddots & \vdots \\
1 				&  0  &  \cdots  &  1  		&  0  &  \cdots  & \chi_{N}  	& C_{N,N+2} & \cdots & C_{N,M} 
 \end{pmatrix} \; .
\label{eq:cola}
\end{align}
This increases the number of inputs from $N$ to $M$, where $M$ is a
tunable hyper-parameter of the network. The entry $\chi_j$ is new for
the quark--gluon implementation and encodes the information if a
particles is charged or not, $\chi_j = 0,1$~\cite{in_color}. For most
of the phase space considered in this paper, we will find that the tagging
performance for our specific applications hardly improves, but
obviously this result should not be generalized.  To make it easier
for the network to learn the mathematical structure of Lorentz
transformations we pass the CoLa output to a Lorentz layer (LoLa)
\begin{align}
\tilde{k}_j 
\stackrel{\text{LoLa}}{\longrightarrow}
\hat{k}_j = 
 \begin{pmatrix*}[c]
  m^2(\tilde k_j)\\ 
  p_T(\tilde k_j)\\ 
  p_T(\tilde k_j) \Delta R_{j,\text{jet}} \\[2mm]
  w^{(E)}_{jm} \,E(\tilde k_m)\\ 
  w^{(d)}_{jm} \, d^2_{jm}\\ 
  E_T(\tilde k_j) E_T (\tilde k_m)  (\Delta R_{jm})^{0.2} \\
 \end{pmatrix*} \; ,
\label{eq:lola}
\end{align}
with $d^2_{jm} = (\tilde k_j - \tilde k_m)^2$. To adapt this layer to
quark--gluon separation we augment it with the third and the last
entries. They follow the definition of the the subjet variables
$w_\text{PF}$ and $C$ in Eq.\eqref{eq:qg_obs}, with the sum over
constituents stripped off so that they are defined per constituent.
The first three $\hat{k}_j$ map individual 4-momenta $\tilde{k}_j$
onto their invariant mass and transverse momentum. The fourth entry is
a linear combination of all energies with trainable weights
$w^{(E)}_{jm}$, while the fifth entry sums over the Minkowski distance
between $\tilde k_j$ and all other 4-momenta $\tilde k_m$, again
weighted by $w^{(d)}_{jm}$ which is updated after each training
epoch. For the lower three entries we can either sum over or minimize
over $m$ while keeping $j$ fixed. For $w^{(E)}_{jm}$ we choose the sum
over the internal index; for $w^{(d)}_{jm}$ we include four copies
with independently trainable weights, two summing and two minimizing
over the internal index; for the last entry we use two copies, one
with a sum and one with a minimum.  However, it turns out that the new
\textsc{LoLa} observables have limited impact on the quark--gluon
separation, independent of the options applied to the last the last
entry in Eq.\eqref{eq:lola}.

\begin{figure}[t]
\includegraphics[width=0.61\textwidth]{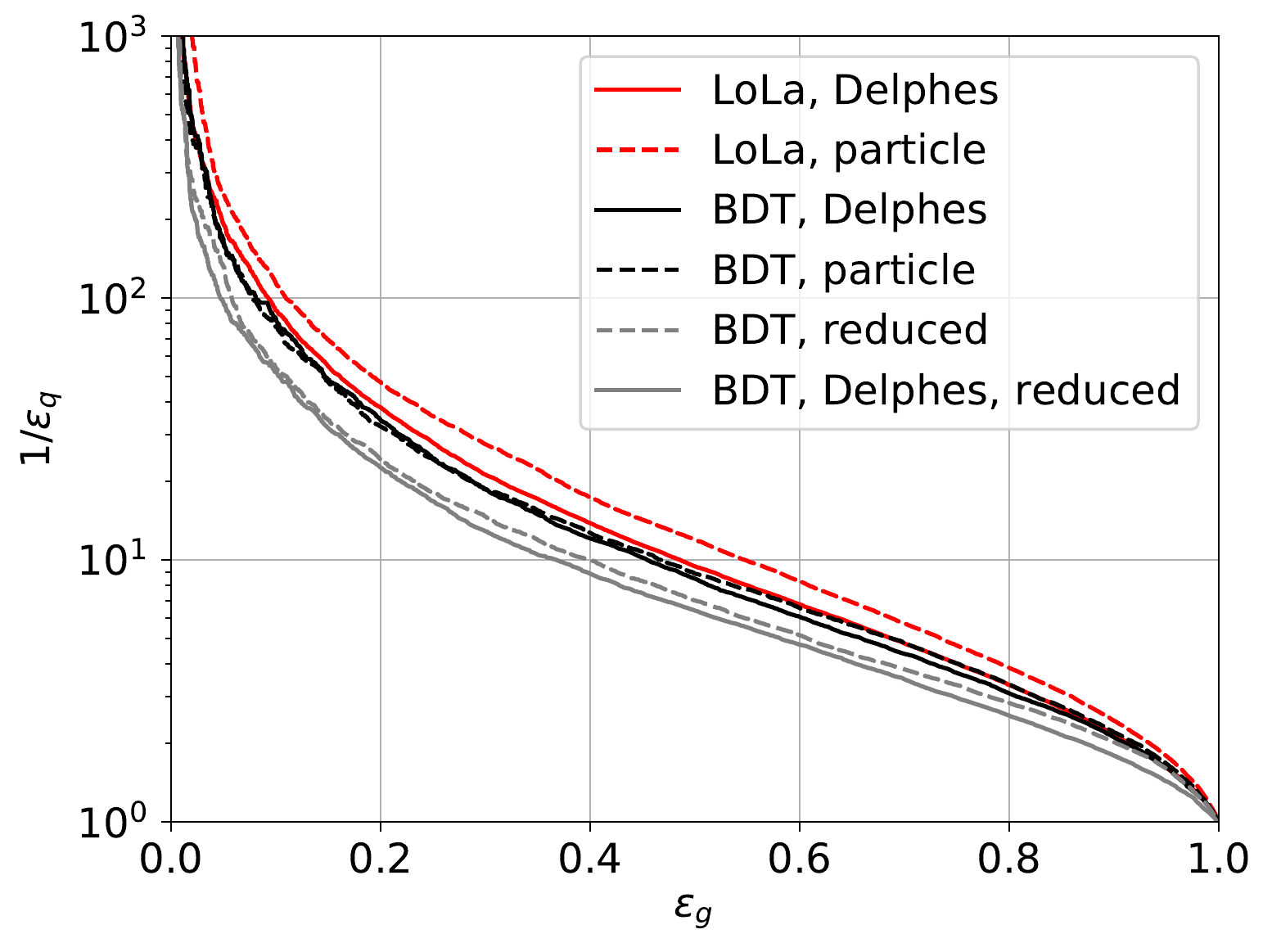}
\hspace*{0.05\textwidth}
\includegraphics[width=0.33\textwidth]{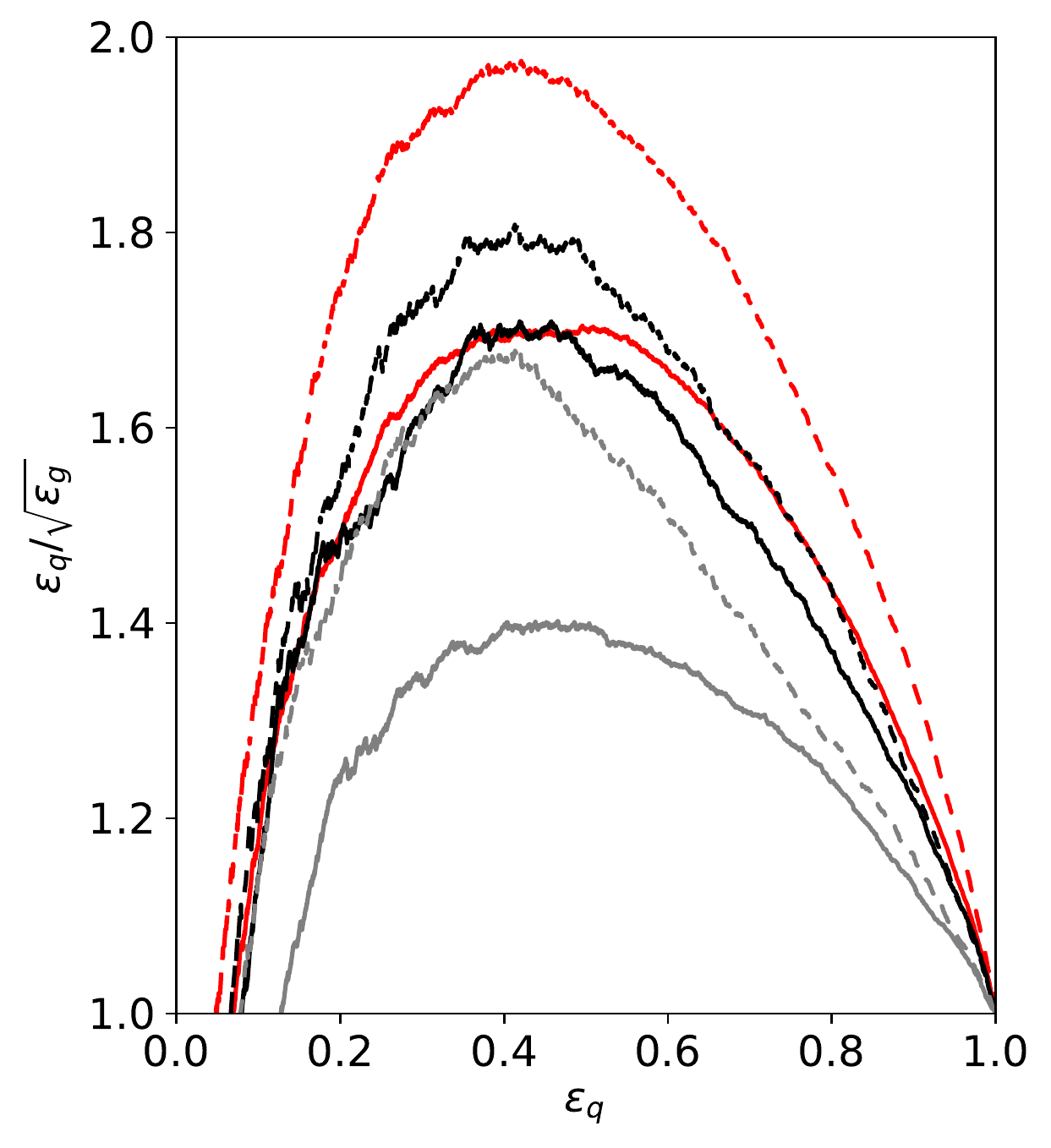}
\caption{ROC and significance improvement curves for the \textsc{LoLa}
  tagger trained and tested on pure samples, with and without detector
  simulation. We compare to a BDT analysis of the full set of of
  observables, Eq.\eqref{eq:full_set}, and to a reduced set of
  observables, Eq.\eqref{eq:reduced_set}.}
\label{fig:roc_pure}
\end{figure}

After the \textsc{LoLa} stage, the inputs are passed through
ReLU-activated dense layers with 100 and 50 units and dropout rate 0.2
and 0.1, respectively.  Both dense layers have an additional L2
regularization of $5\times10^{-4}$ and are initialized with He-normal
functions. A final dense layer converts the weights into a normalized
score with SoftMax activation. All training is done using
\textsc{Keras}~\cite{keras} with the \textsc{Theano}~\cite{theano}
back-end on a GPU cluster. The hyper-parameters are optimized with
\textsc{Adam}~\cite{adam}, using a learning rate of $10^{-5}$ and a
batch size of 128. We have checked that both, for the size of the
training sample and for the number of constituents our performance
reaches safe plateaus.  Throughout this paper we use $N=80$
constituents, significantly above where we would expect the soft
activity to be universal.

\begin{figure}[t]
\includegraphics[width=0.5\textwidth]{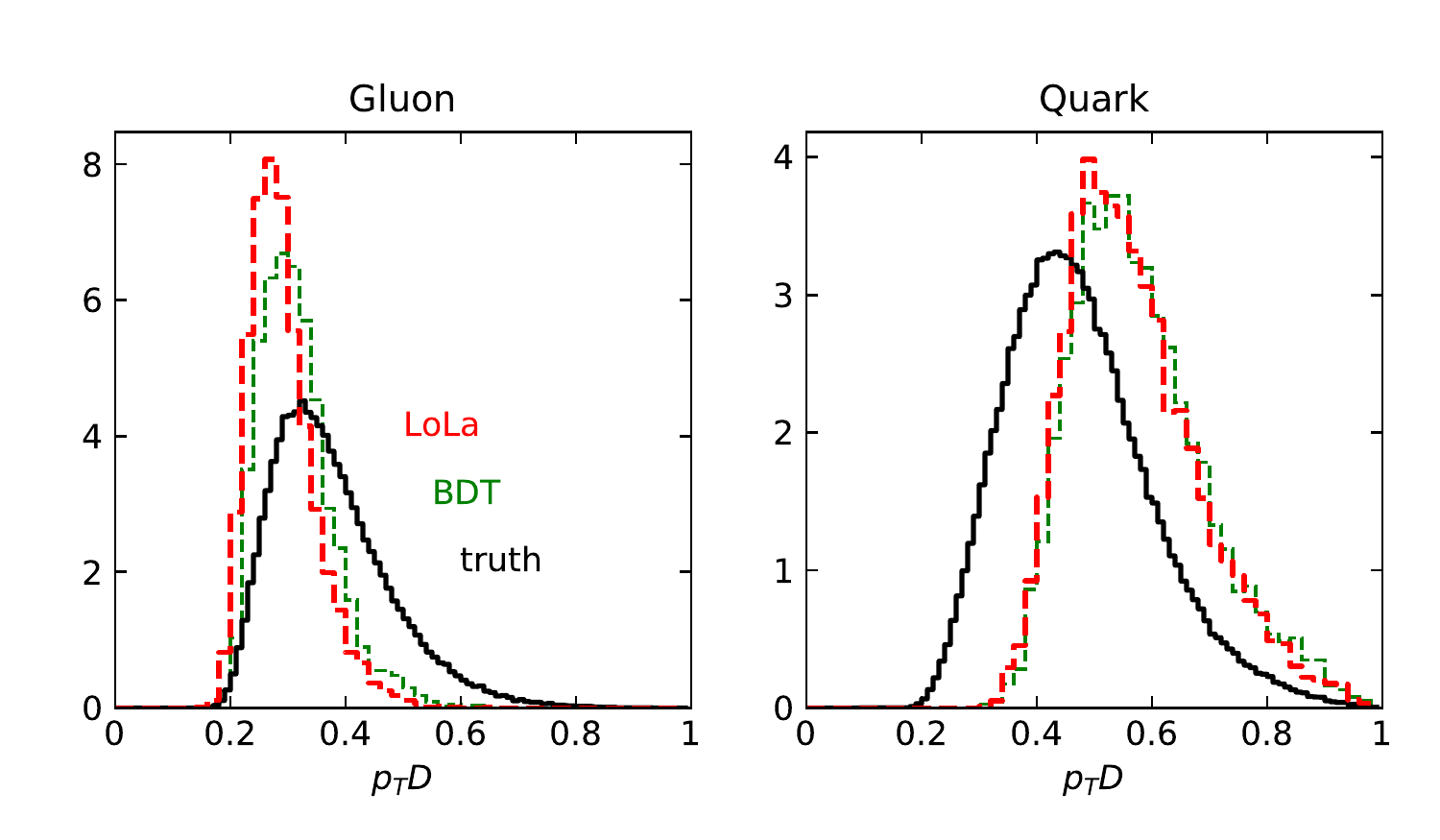}
\includegraphics[width=0.5\textwidth]{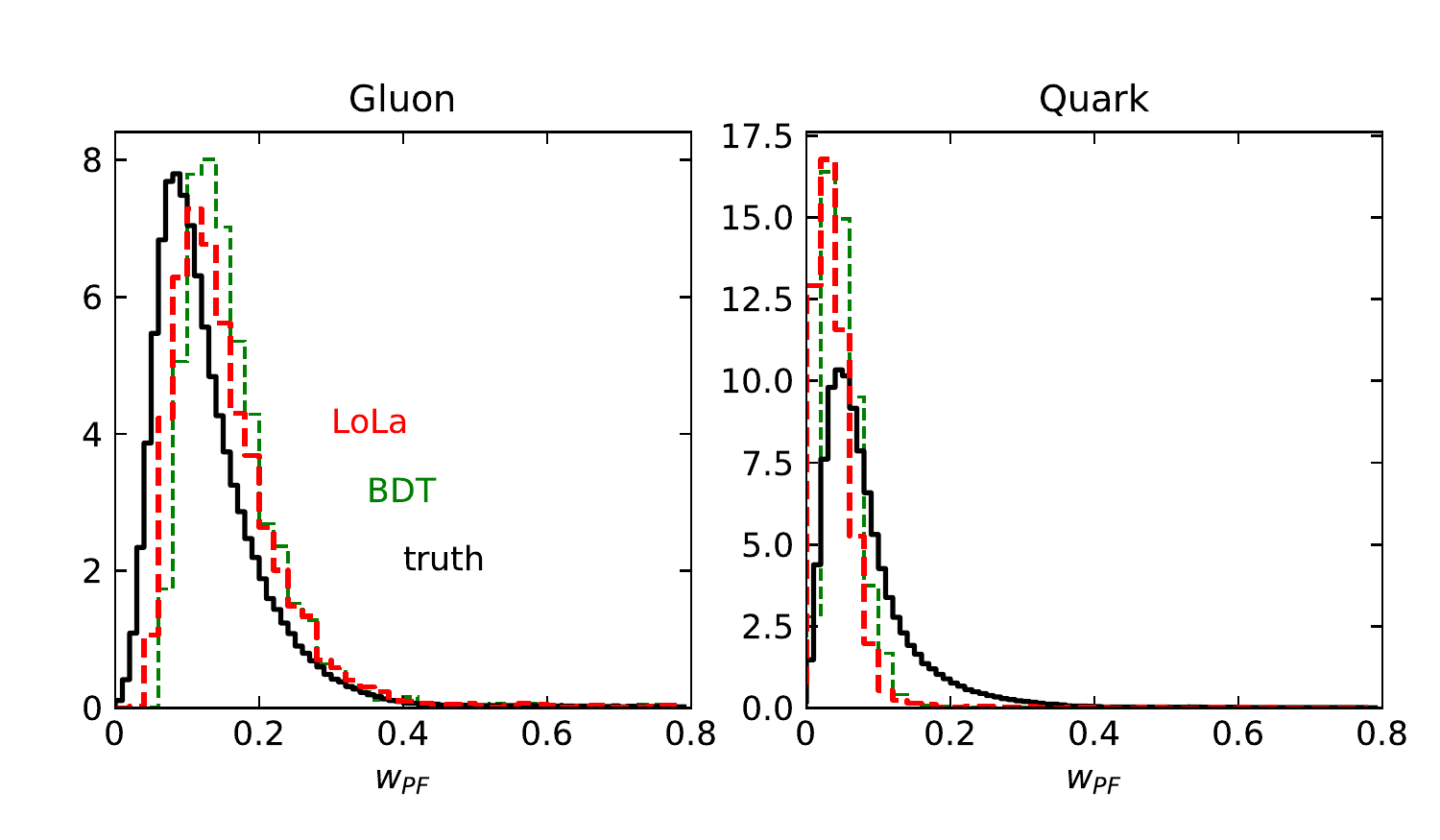}\\
\includegraphics[width=0.5\textwidth]{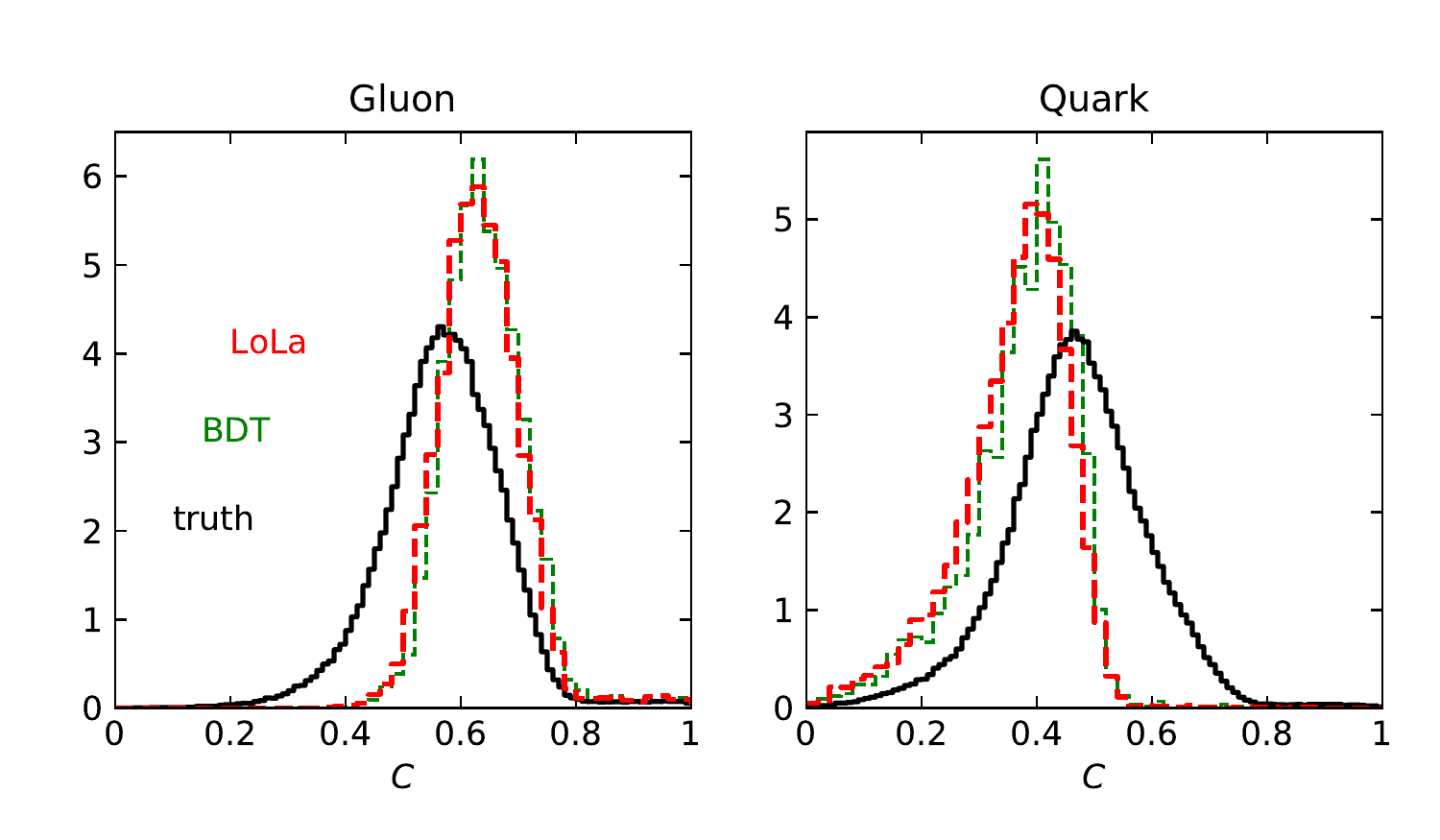}
\includegraphics[width=0.5\textwidth]{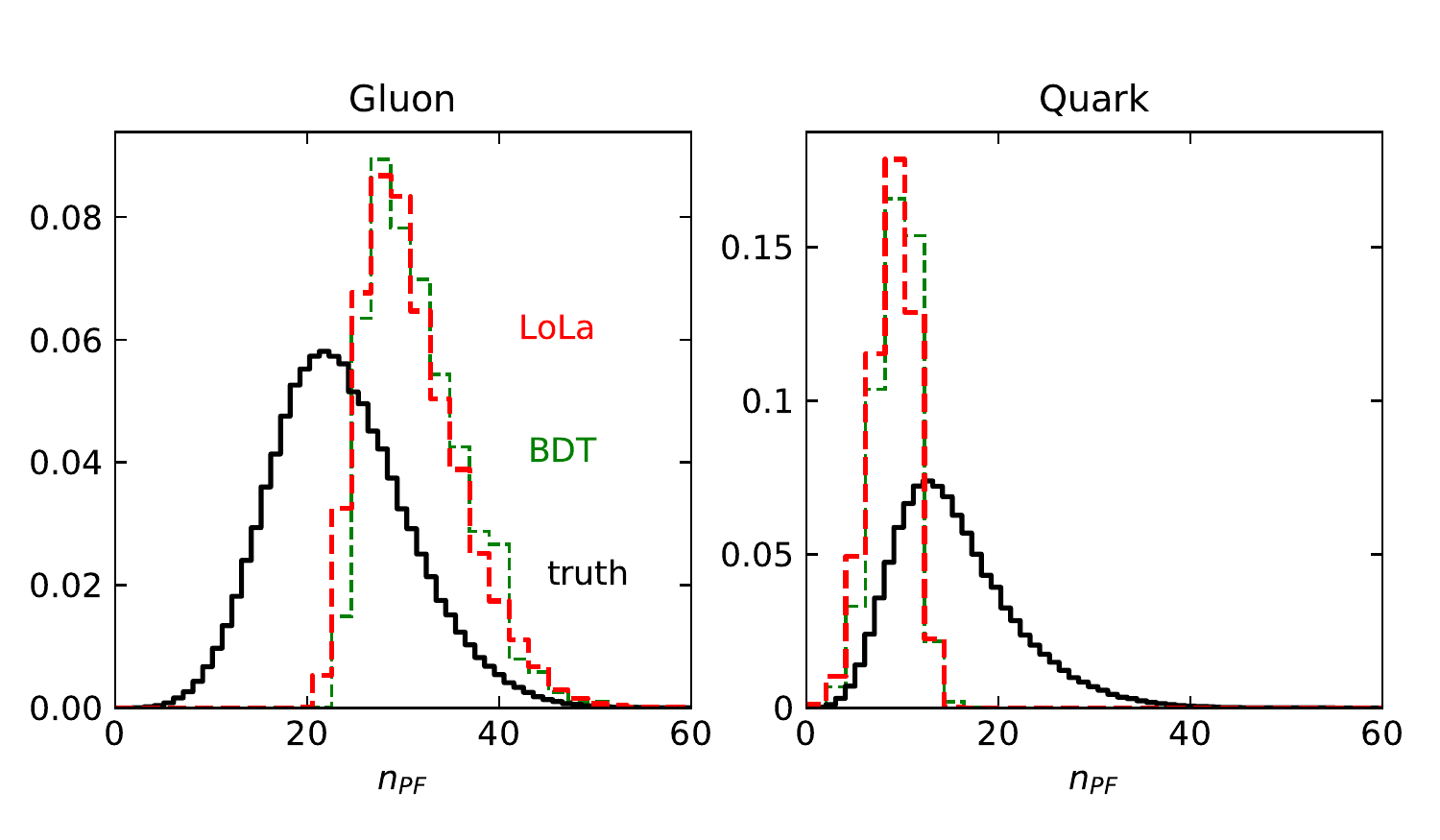}\\
\includegraphics[width=0.5\textwidth]{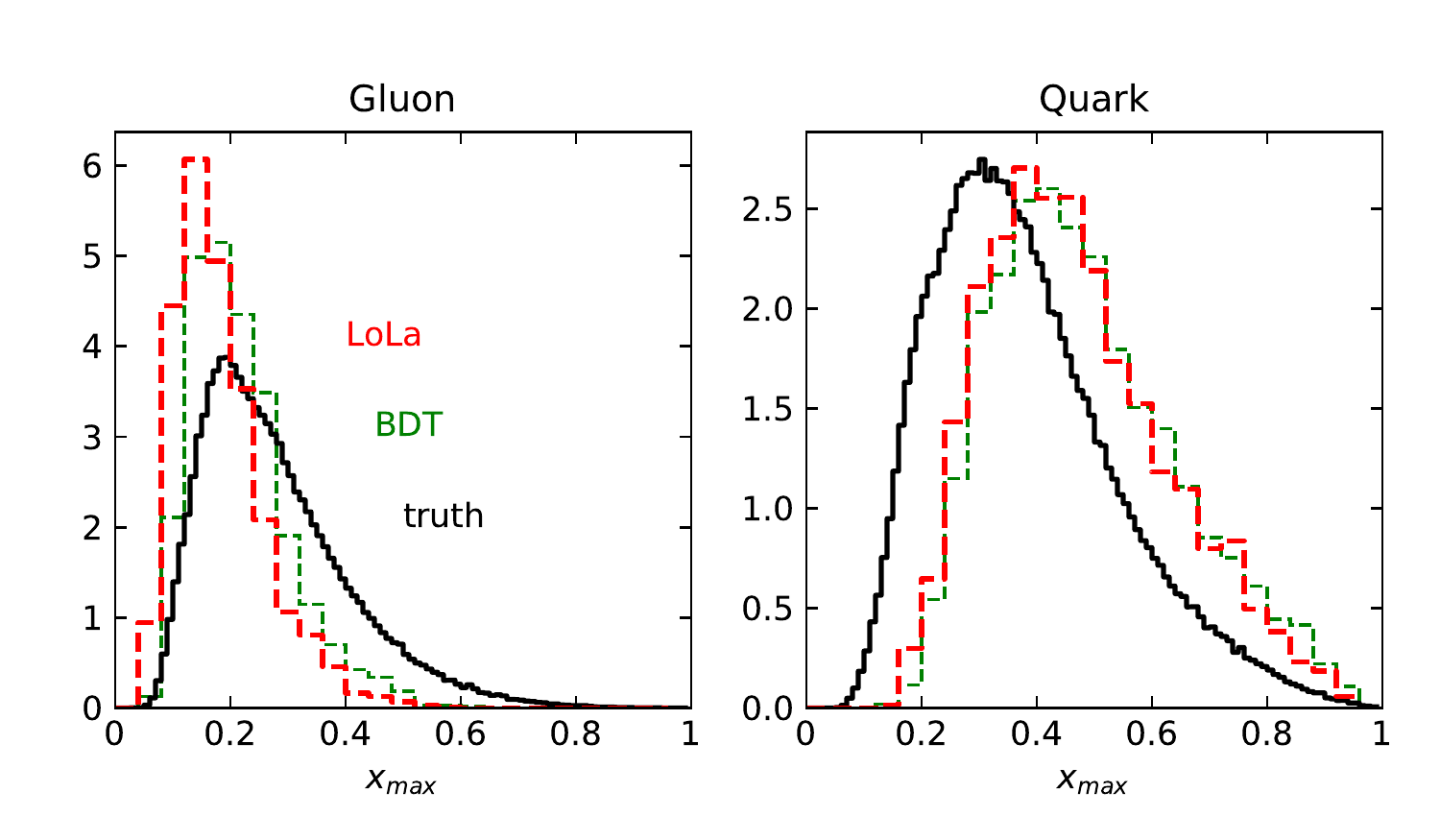}
\includegraphics[width=0.5\textwidth]{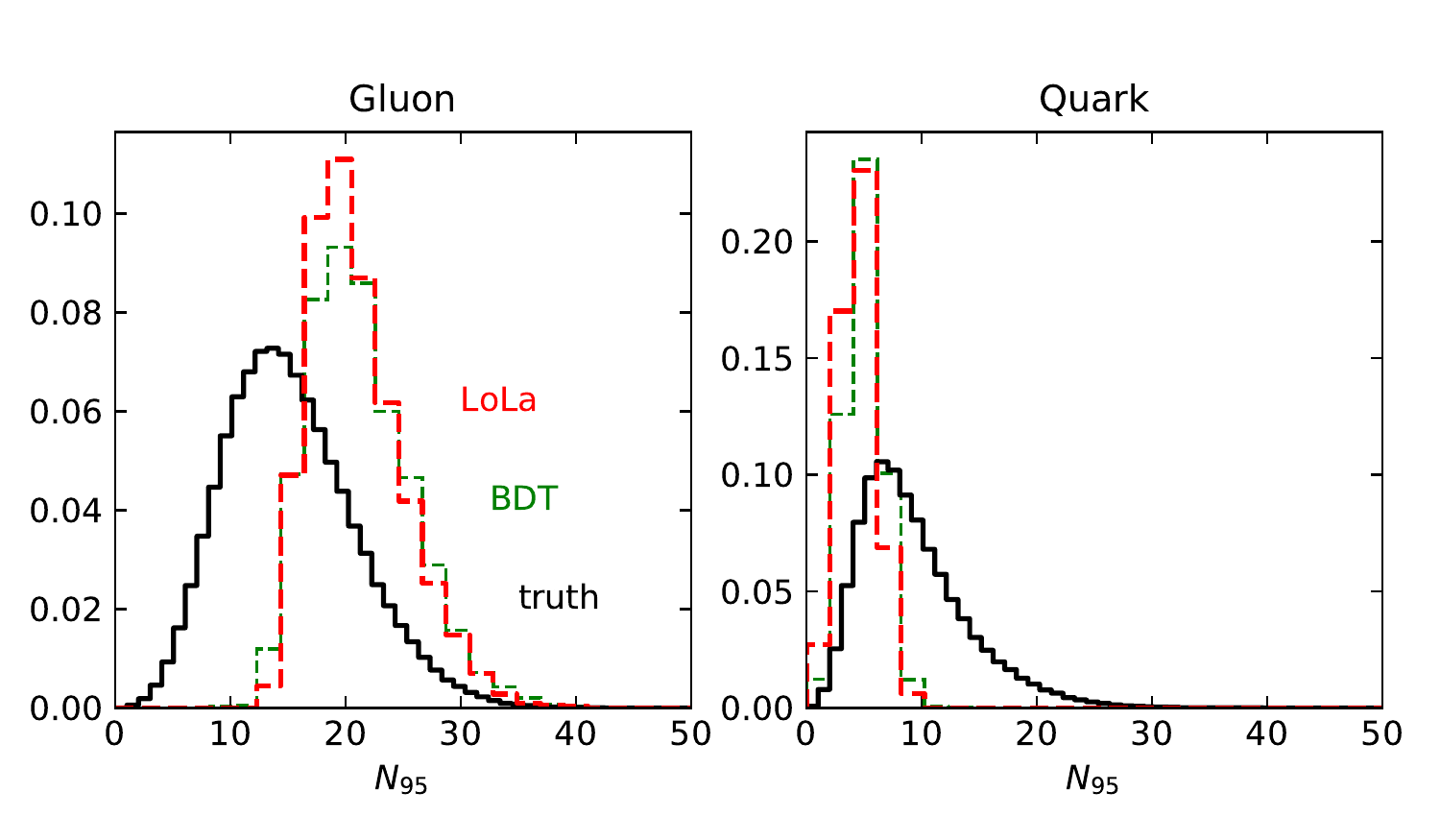}
\caption{Distributions for the sub-jet observables. The black curves
  show the truth from Fig.~\ref{fig:subjet_vars}. The red dotted
  curves are the 30\% most gluon-like or quark-like jets from
  \textsc{LoLa}, the green curves from the BDT.}
\label{fig:cut_plots}
\end{figure}

Turning to the performance, we plot the ROC curves for our
best-performing \textsc{LoLa} architecture in the left panel of
Fig.~\ref{fig:roc_pure}, compared to the 6-observable BDT,
\begin{align}
\left\{ \, 
n_\text{PF}, w_\text{PF}, p_TD, C_{0.2}, x_\text{max}, N_{95} 
\, \right\} \; .
\label{eq:full_set}
\end{align}
In the right panel we also show the increase in signal significance as
a function of the signal efficiency, to help us optimize the impact of
the tagger as an analysis tool for instance in terms of $\text{SI} =
\epsilon_g/\sqrt{\epsilon_q}$.  With our \textsc{LoLa} network we
reproduce the performance of the enhanced images setup of
Ref.~\cite{in_color} without detector simulation and after accounting
for the move from \textsc{Pythia} to \textsc{Sherpa}. Our agreement is
at the level that different trainings of our \textsc{LoLa} tagger on
the framework of Ref.~\cite{in_color} shower a stronger variation than
the agreement between the \textsc{LoLa} and the CNN performance.
Different architectures without detector effects are studied in detail
in Ref.~\cite{deep_sets}. They are very close in performance,
including convolutional networks like that of Ref.~\cite{in_color},
and we have good reason to assume that this pattern will not change
once we include detector effects.

We also note an overall improvement with respect to our 6-observable
BDT.  The fact that the deep network does not hugely outperform the
multi-variate analysis on the subjet level is not unexpected.  The
difference between the \textsc{LoLa} network and the BDT becomes
smaller once we include detector effects. This points to the deep
network finding additional information which even the theoretically
poorly defined observables do not capture. As a test of stability we
also show BDT results with a reduced and less IR-sensitive set of
observables,
\begin{align}
\left\{ \, 
p_TD, C_{0.2}, x_\text{max}, N_{95} 
\, \right\} \; .
\label{eq:reduced_set}
\end{align}
As we can see in Fig.~\ref{fig:roc_pure} this reduces the over-all
performance of the BDT, but does not improve the stability with
respect to detector effects.\medskip

Finally, we need to test if the quark--gluon network correctly
captures the information we know exists at the subjet
level~\cite{what}.  Because we have access to Monte Carlo truth we
can, for instance, plot the distributions of our six observables for
quark jets identified as quarks and for gluon jets identified as
gluons. We can compare these distributions between the \textsc{LoLa}
network, the BDT, and the truth information, all including detector
effects.  In Fig.~\ref{fig:cut_plots} we plot all observables
introduced in Sec.~\ref{sec:obs}, at truth-level and after selecting
the 30\% best-identified jets.  For gluon jets the classifier favors
slightly lower values of $p_TD$ and $x_\text{max}$, and larger values
of $C$, $N_{95}$ and $n_\text{PF}$. A significant sculpting of these
distributions relative to truth indicates a challenge in separating
the two hypotheses. The observables where \textsc{LoLa} best matches
the truth are $w_\text{PF}$ and $C_{0.2}$. These are also the two most
important observables in the BDT in Fig.~\ref{fig:bdt_rocs},
indicating that the BDT and \textsc{LoLa} indeed rely on similar
information.

\subsection{Jet kinematics}
\label{sec:jets}

\begin{table}[b!]
\begin{footnotesize} 
\begin{minipage}[t]{0.49\textwidth}
\begin{tabular}{c|ccccc}
\toprule
Train & \multicolumn{5}{c}{Test} \\ 
\midrule
& 200-210  &  210-220  &  220-230  &  230-240  &  240-250  \\
\midrule
200-210  &   0.812   &   0.812   &   0.812   &  0.818 &   0.816   \\   
210-220  &   0.812   &   0.813   &   0.812   & 0.819   &   0.817   \\   
220-230  &  0.804   &   0.805   &   0.810   &  0.811   &   0.808   \\   
230-240  &   0.803   &   0.804   &   0.801   &  0.814   &   0.809   \\   
240-250  &   0.810   &   0.811   &   0.811   & 0.820    &   0.818   \\
\bottomrule
\end{tabular} 
\end{minipage}
\hspace*{0\textwidth}
\begin{minipage}[t]{0.49\textwidth}
\begin{tabular}{c|ccccc}
\toprule
Train & \multicolumn{5}{c}{Test} \\ 
\midrule
&  200-250  &  300-350  &  400-450  &  500 - 550  &  600-650  \\
\midrule
  200-250  &   0.813   &   0.818   &  0.805  &  0.782  &    0.74  \\  
  300-350  &   0.811   &   0.825   &  0.823  &  0.818  & 0.80 \\   
  400-450  &   0.809   &   0.824   &  0.834  &  0.838  & 0.80 \\   
  600-650  &   0.807   &   0.816   &  0.830  &  0.840  & 0.841   \\   
\bottomrule
\end{tabular} 
\end{minipage}
\end{footnotesize}
\caption{Areas under the ROC curve for the \textsc{LoLa} tagger
  trained and tested on pure samples sliced in $p_{T,j}$. The
  uncertainty on each entry is one to two units on the last shown
  digit.}
\label{tab:pure}
\end{table}

One dangerous sources of systematic uncertainties in subjet physics
and elsewhere is mis-measuring the momentum of the
jet~\cite{wjets}. Because the structure of parton splittings is
sensitive to the range of energies described by the splitting history,
we do not want to remove this information for example through an
adversarial network. Instead, we want to include $p_{T,j}$ in the
information available to the tagger. Before we do so, we need to
understand at what level the quark--gluon network is sensitive to the
transverse momentum of the jet~\cite{in_color,qg_rnn}.

To this end we train and test individual \textsc{LoLa} networks in
different slices of $p_{T,j}$, again with detector effects, and test
them on over a range of transverse momenta.  We show the AUC values
for different combinations of training and testing samples in
Tab.~\ref{tab:pure}.  The left table shows the performance of the
network for a small step size $\Delta p_T = 10$~GeV. On the diagonal
we see that the performance of the network slightly increases towards
higher momenta. This can be understood through the larger number of
constituents radiated off initial partons with higher momentum. For
the off-diagonal entries there is also a small generic trend that
using a network on somewhat higher-$p_T$ jets than it was trained for
does not reduce its efficiency. Because the differences between quarks
and gluons are more subtle for softer jets, a network trained on these
subtle differences may also be applied to harder jets. However, in the
other direction the network trained on the more obvious hard jets will
slightly deteriorate for softer jets. In the right table we test a
wider range of transverse momenta. We observe the same trend, but for
networks trained between 200 and 350~GeV the performance seriously
suffers when we compare it to $p_T > 600$~GeV. 

We only show central values in both of these tables, but we have
estimated uncertainties on the performance measures in two ways. The
larger error bar comes from using a trained network on different test
samples, it gives typical uncertainties of $\Delta \text{AUC} \approx
0.002$ for most of the entries, increasing to $\Delta \text{AUC}
\approx 0.01$ for the larger separations in $p_T$. The error we find
from using different trainings on the same test sample is, in our
case, about an order of magnitude smaller.

\begin{figure}[t]
\centering
\includegraphics[width=0.47\textwidth]{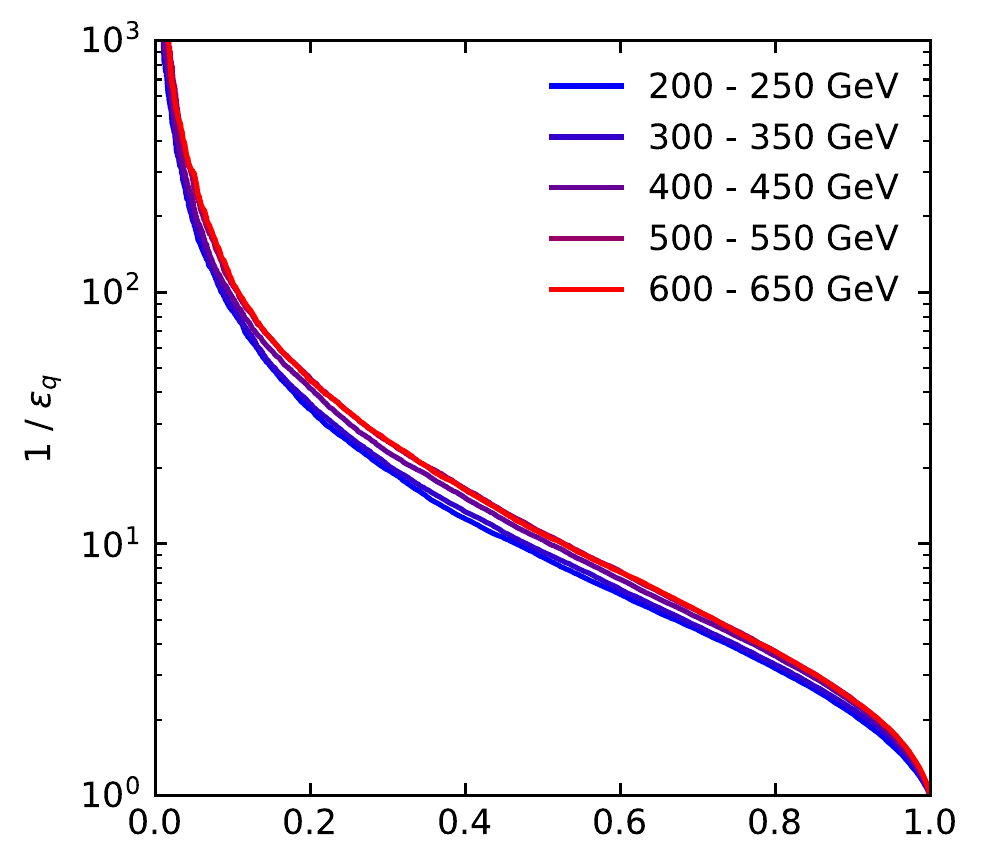}
\hspace*{0.01\textwidth}
\includegraphics[width=0.47\textwidth]{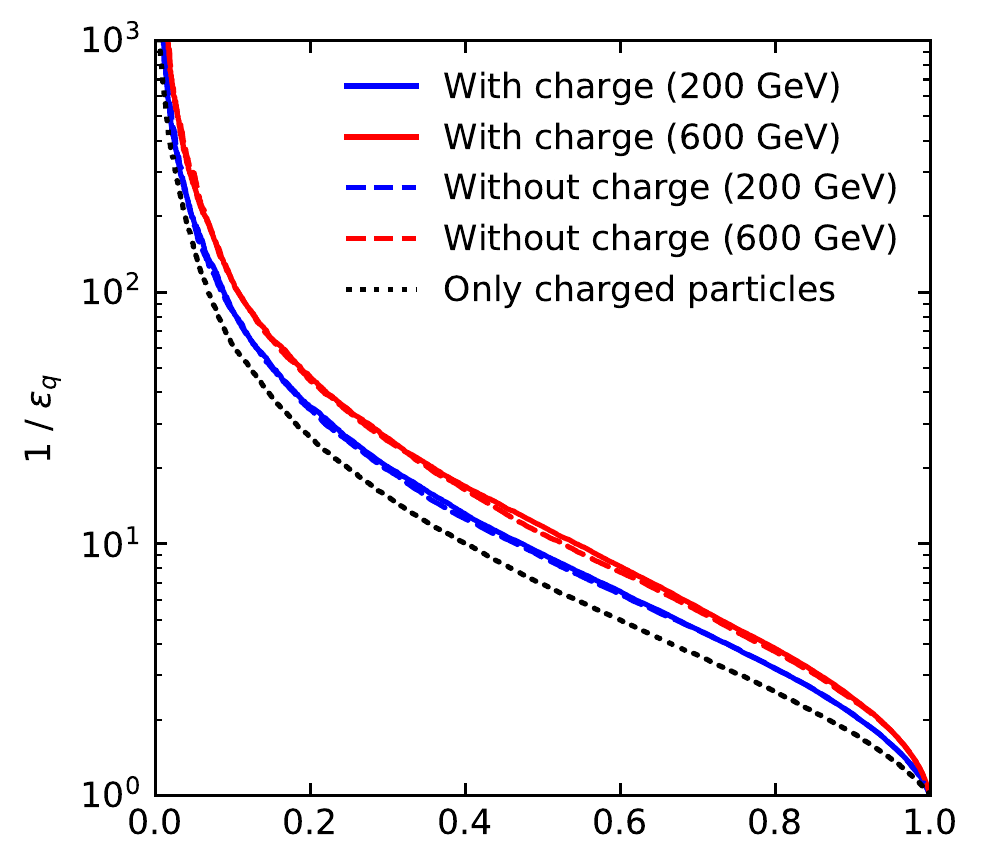}
\hspace*{0.002\textwidth}
\includegraphics[width=0.47\textwidth]{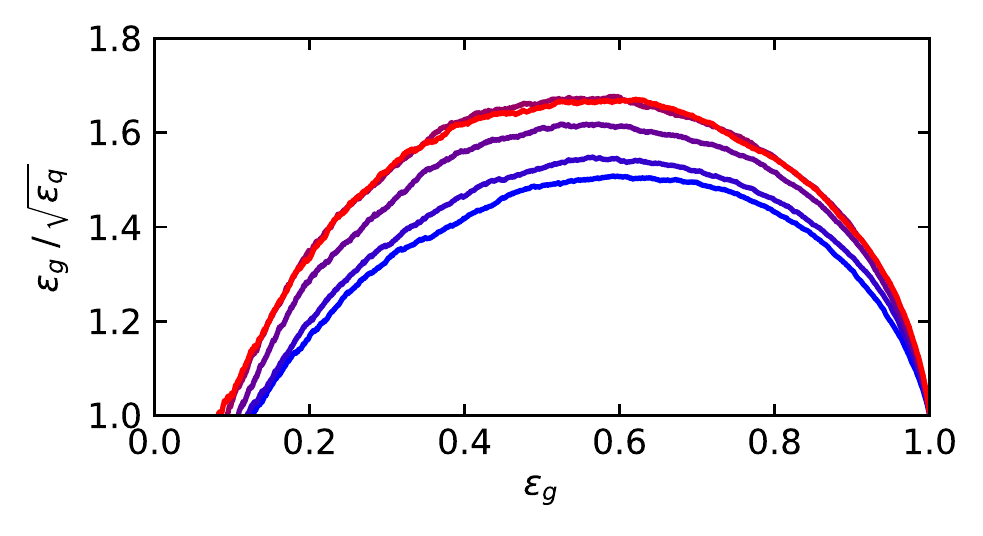}
\hspace*{0.01\textwidth}
\includegraphics[width=0.47\textwidth]{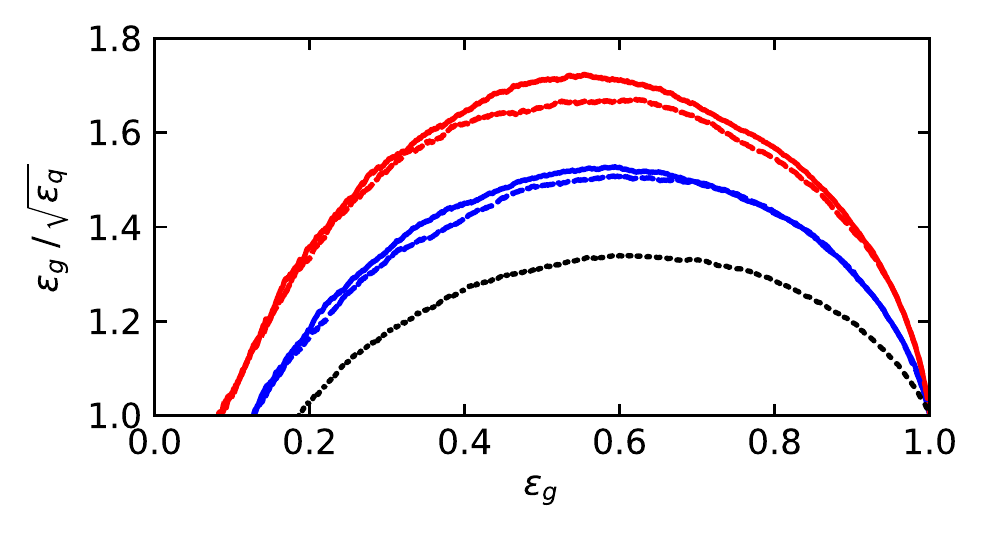}
\vspace*{-6mm}
\caption{Left: ROC and SI curves for the pure quark and gluon samples
  in non-overlapping jet $p_T$ ranges. Right: ROC and SI curves for
  the pure quark gluon samples including charge information.}
\label{fig:pure_final}
\end{figure}

For the $p_{T,j}$ slices in Tab.~\ref{tab:pure} we can compute the ROC
curves for the \textsc{LoLa} quark--gluon discrimination. In the left
panel of Fig.~\ref{fig:pure_final} we see how the performance of the
tagger is stable, with a slight increase in performance towards higher
jet momenta.

In the right panel of Fig.~\ref{fig:pure_final} we repeat the same
exercise, but including the charge information discussed in
Eq.\eqref{eq:cola}. Indeed, the performance is unchanged for this
specific change in the \textsc{LoLa} setup, at least up to $p_{T,j} <
600$~GeV and once we include detector effects.

\section{Mono-jets} 
\label{sec:mono}

To see at what level quark--gluon discrimination really helps at the
LHC we need benchmark applications. For WBF jets we have unfortunately
seen that the substructure of the tagging jets can alleviate the
pressure on global observables like a central jet veto, but that the
signal vs background system is already over-constrained by event-level
kinematic information and jet substructure~\cite{onke}. We therefore
turn to the simplest jet analyses with the fewest number of
established handles to control the background.

Our first candidate is the mono-jet signature probing invisible decays
of a SM-like Higgs boson. Here, the transverse momentum of the tagging
jet is essentially the only kinematic variable used in standard
analyses.  Far from the expected performance of the leading WBF and
$VH$ channels for invisible SM-like Higgs decays, this mono-jet
channel is extremely versatile when we search for dark matter or want
to learn more about the nature of an invisible Higgs signal. For a
Higgs-like mediator it provides us with a benchmark process for a
tagger extracting a gluon-dominated signal from a quark-dominated
background~\cite{monojet_qg}. Obviously, all our findings can be
generalized to searches for (pseudo-)scalar mediators at the LHC. For
those the relative importance of the electroweak WBF and $VH$ channels
compared to the gluon-induced mono-jet channel can obviously be
completely altered.

\begin{figure}[t]
\begin{center}
\begin{minipage}[t]{0.34\textwidth}
 \includegraphics[width=0.80\textwidth]{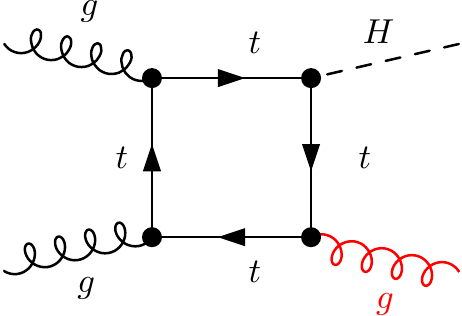} \\
 \includegraphics[width=0.80\textwidth]{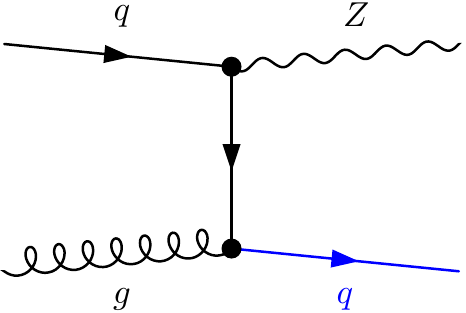}
\end{minipage}
\begin{minipage}[l]{0.65\textwidth} 
\includegraphics[width=0.95\textwidth]{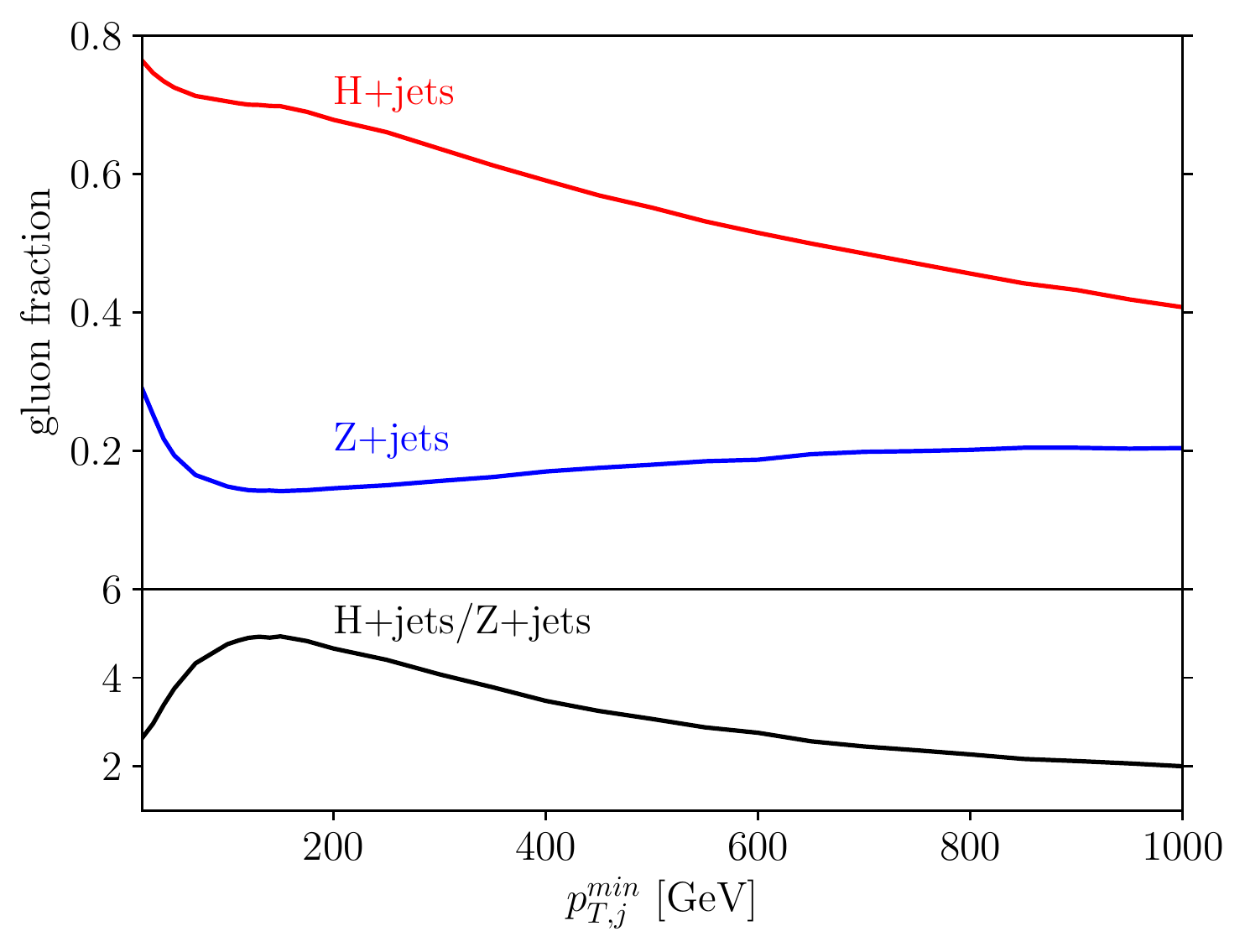}
\end{minipage}
\end{center}
\caption{Left: representative Feynman diagrams for the mono-jet signal
  and $Z$+jets background. Right: fraction of events with a leading
  jet above $p_{T,j}^\text{min}$ with gluon-initiated leading jets in
$H$+jets events (red) and quark-initiated jets in $Z$+jets events
(blue) as a function of $p_{T,j}$. The bottom panel shows the ratio of
the two gluon fractions.}
\label{fig:feynman}
\end{figure}

The key feature of mono-jet searches with scalar mediators is that the
signal jet is almost always gluon-initiated, while for the $Z$+jets
background it is mostly quark-initiated, as illustrated in
Fig.~\ref{fig:feynman}.  Increasing $p_{Tj}$ pushes the events
kinematics towards larger proton momentum fractions and enhances the
quark contribution, slowly reducing the gluon purity of the Higgs
signal.  Observing such a signal in mono-jet events requires exquisite
control of the large backgrounds from $V$+jets production. While the
largest background is $Z(\to\nu\nu)$+jets, there exists a sizeable
irreducible contribution from $W(\to l\nu)$+jets, where the lepton
either fakes a jet or escapes undetected~\cite{vjets}. Due to the rather inclusive
signature of a high-$p_T$ jet with large missing transverse energy,
there is little to cut on other than either $p_{T,j}$ of $\met$. In
practice, a cut of at least $\met \ge 100~\gev$ is typically required
at the trigger level.

\begin{figure}[t]
\centering
\includegraphics[width=0.32\textwidth]{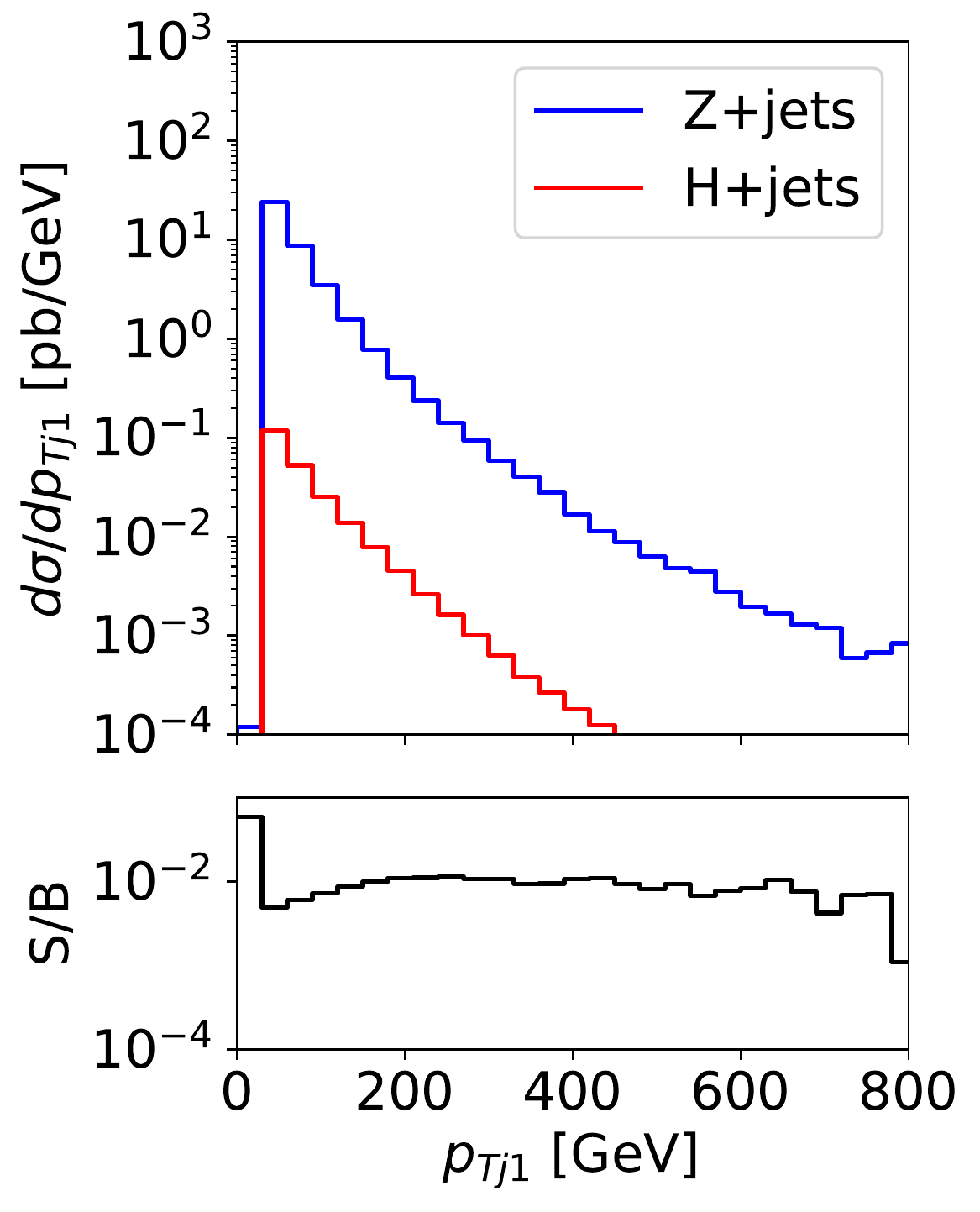}
\includegraphics[width=0.32\textwidth]{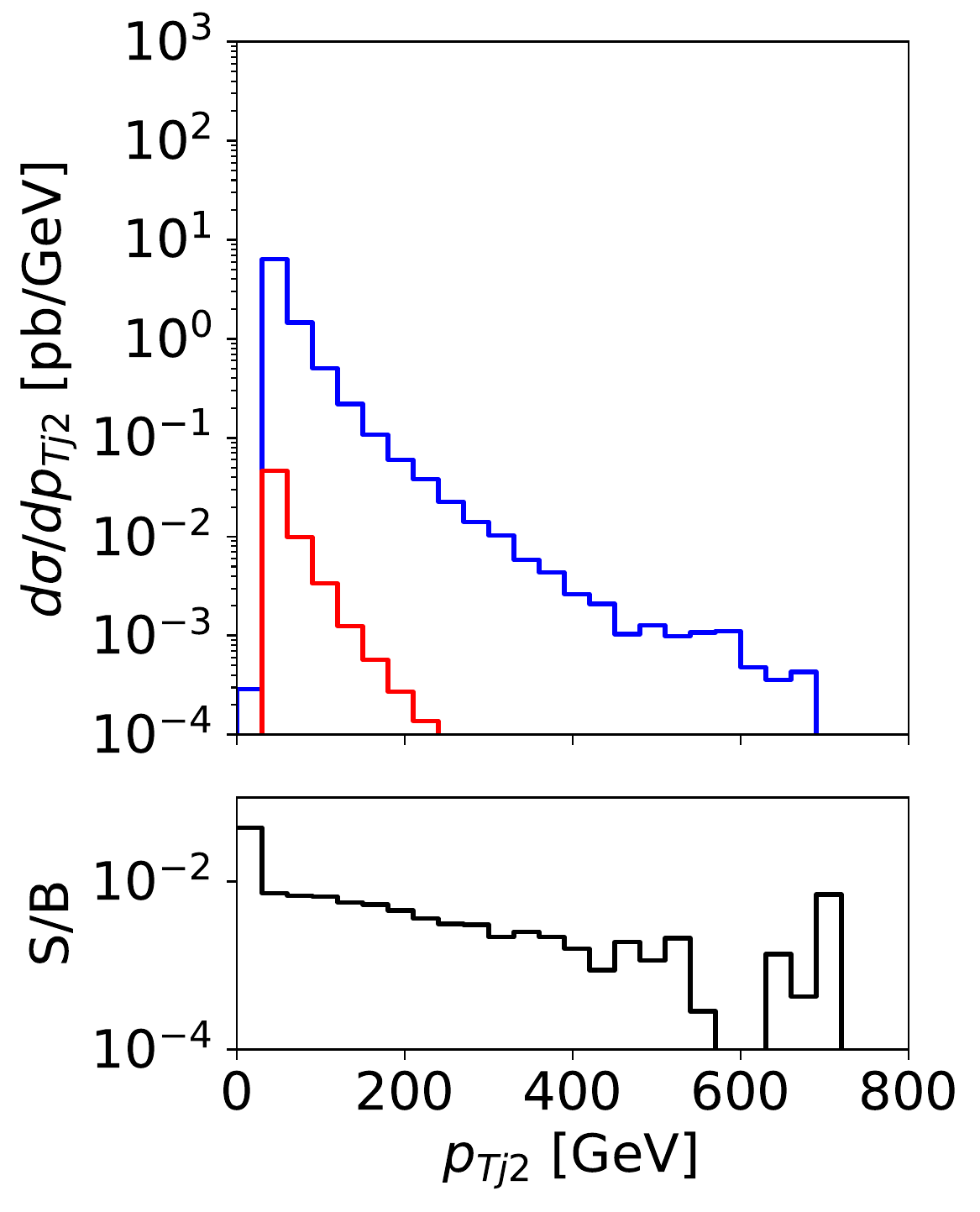}
\includegraphics[width=0.32\textwidth]{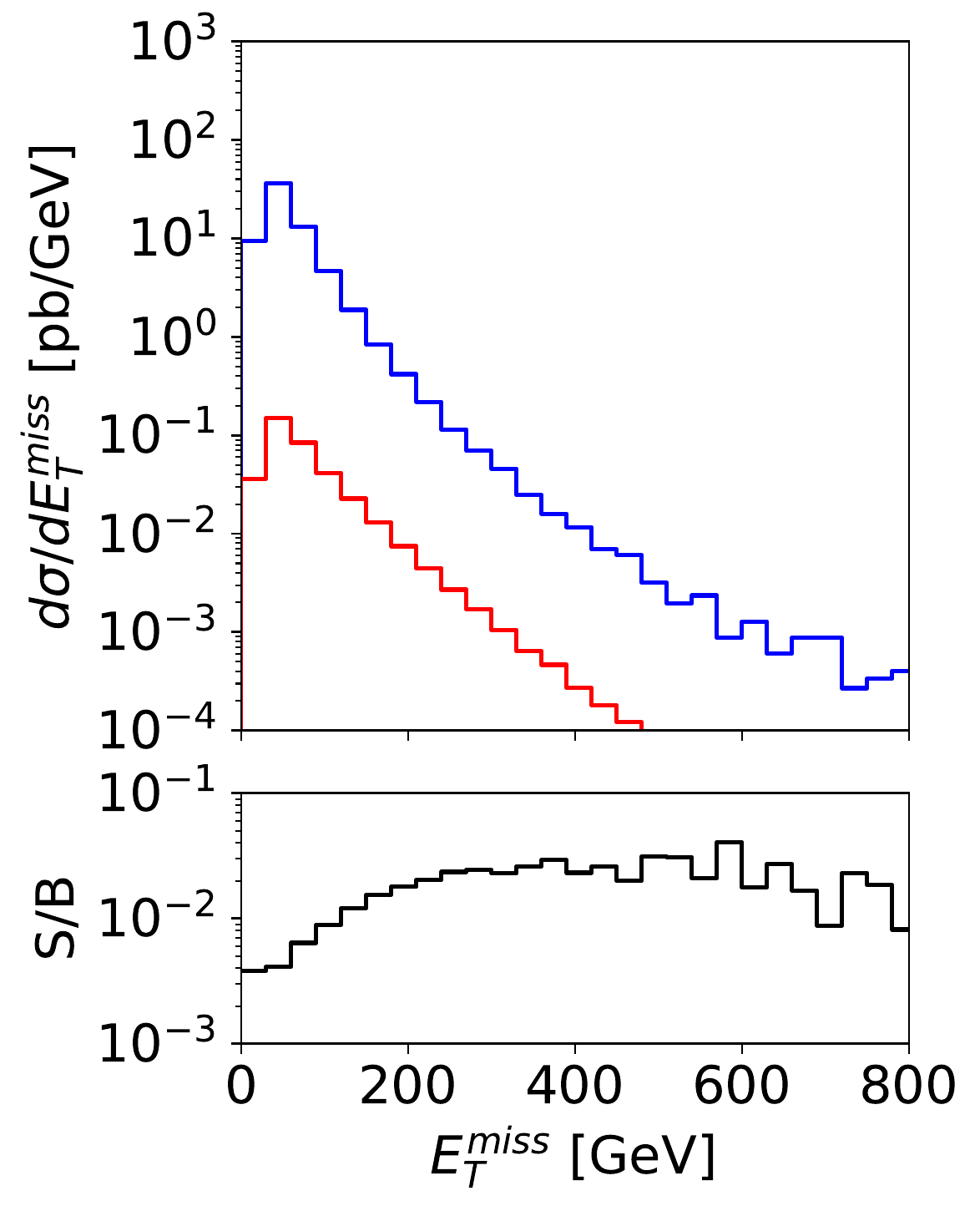}\\
\includegraphics[width=0.32\textwidth]{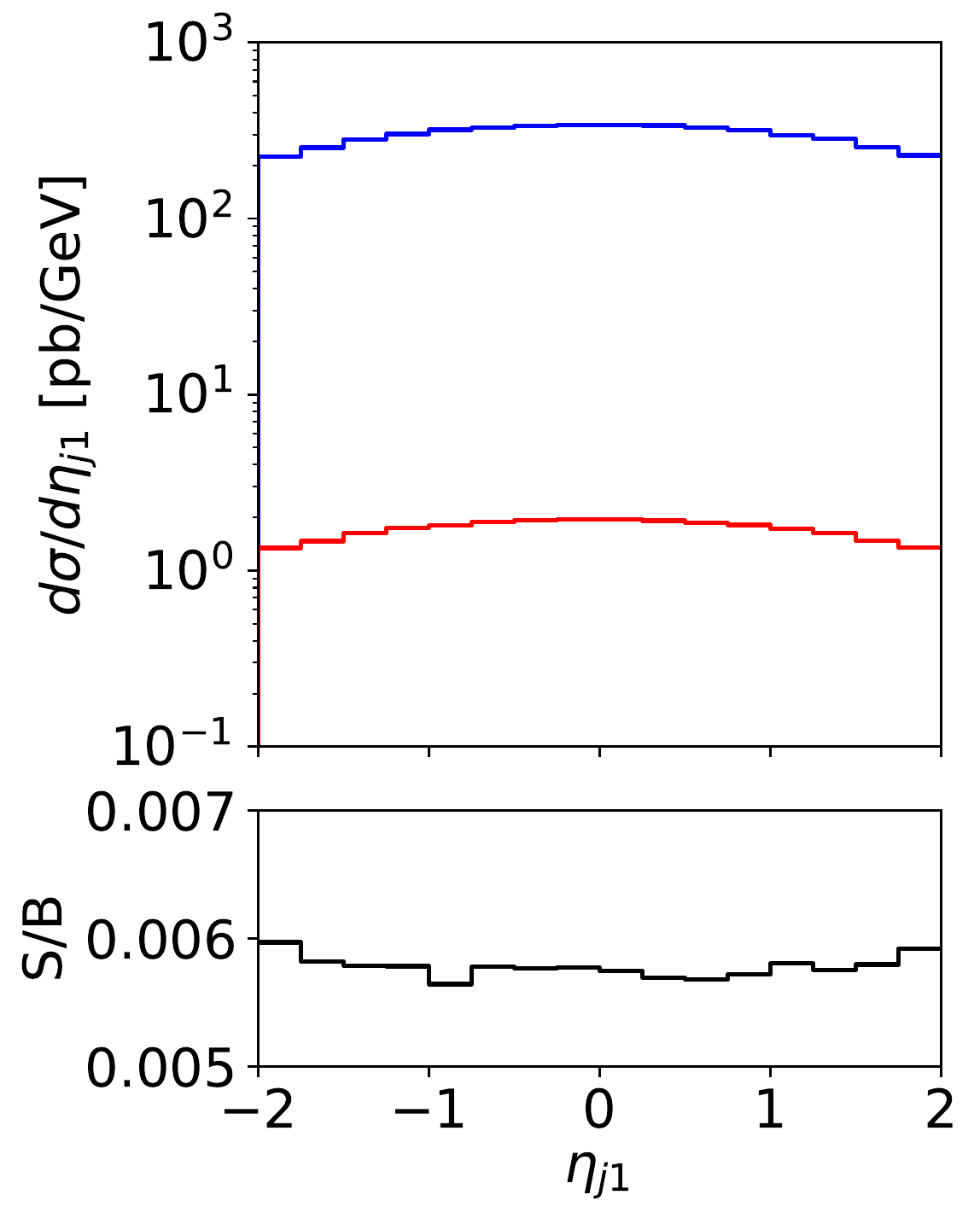}
\includegraphics[width=0.32\textwidth]{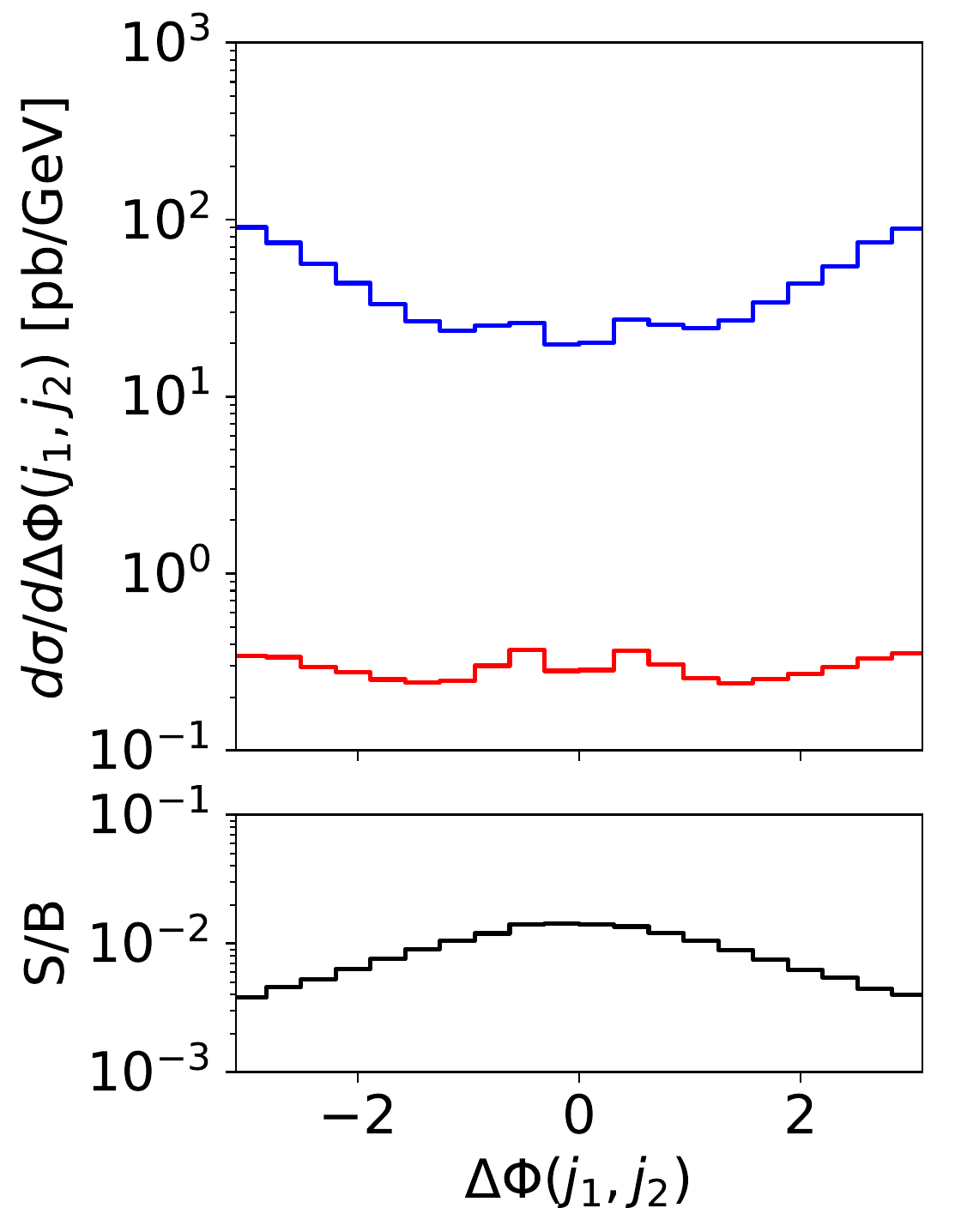}
\includegraphics[width=0.32\textwidth]{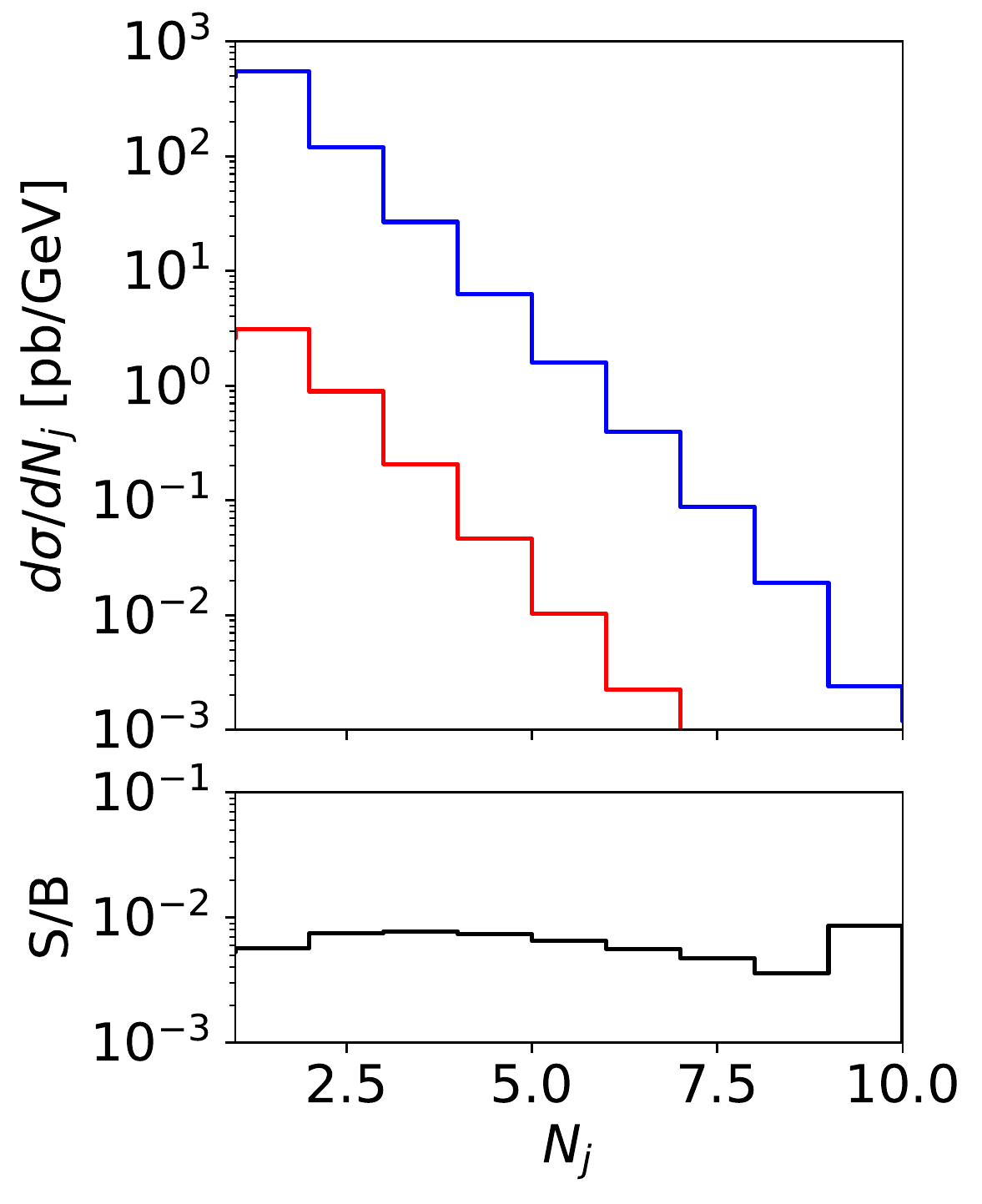}
\caption{Kinematic distributions for the $H$+jets signal (red) and
  leading $Z$+jets background (black), along with the
  signal-to-background ratio.  We show the leading $p_{T,j}$, the
  second $p_{T,j}$, $\met$, the pseudorapidity of the leading jet,
  $\Delta\phi$ of the leading two jet, and the jet multiplicity.}
\label{fig:kin_mono}
\end{figure}

We generate the $H$+jets signal events, including the finite top mass
effects with \textsc{Sherpa}2.2.1~\cite{sherpa} and
\textsc{OpenLoops}~\cite{openloops} at a collider energy of
14~TeV. For the $Z$+jets background we also use
\textsc{Sherpa}2.2.1~\cite{sherpa} with the \textsc{Comix} for matrix
element generation~\cite{comix}, and we employ \textsc{Ckkw-L}
merging~\cite{ckkw} with up to two jets in the matrix element for both
$H$+jets and $Z$+jets. As in the case of the pure samples, we use
$\Delta R=0.4$ anti-$k_T$ jets with all visible final-state particles
of $|\eta| < 2.5$ as constituents~\cite{fastjet}. As long as we stick
to leading-order simulation we can extract the parton content for
example of the hardest jet from Monte Carlo truth.

To illustrate the challenge in observing this signal, we plot some
kinematic distributions for the signal and background in
Fig.~\ref{fig:kin_mono}. Note that following the discussion in
Sec.~\ref{sec:obs} we do not distinguish gluon jets from quark jets,
but the Higgs plus jets signal from the $Z$ plus jets background.
First, the expected signal-to-background ratio even assuming an
invisible Higgs branching ratio of formally 100\% is at the per-mille
level. Second, the leading jet kinematics for the signal and
background is essentially identical, while the second jet is actually
softer in the signal. A cut-and-count analysis above a stringent
$\met$ requirement is not an optimal analysis strategy, because the
small difference between the Higgs and $Z$ masses hardly affects the
kinematics. Of course, if the mono-jet signal is due to a light
mediator, the signal $p_T$-spectrum will be harder.

A subjet feature, which is not exploited in the event-level analysis
is that the hardest background jet is quark-initiated in $80\%$ of
events, while the leading signal jet is usually gluon-initiated.  From
Fig.~\ref{fig:feynman} we expect the quark--gluon tagger to be most
useful at low to intermediate $p_{Tj}$. To study this question
quantitatively, we generate mono-jet samples in non-overlapping
slices of $p_{T,j}$ and train and test \textsc{LoLa} on all
combinations of the above samples. The performance of each
combination, given by the area under the curve (AUC), is shown in
Tab.~\ref{tab:mono}. These numbers can be directly compared to their
counterparts for pure samples in Tab.~\ref{tab:pure}.  We see that the
diagonal entries, corresponding to networks trained and tested in the
same $p_T$ range, show the best performance, and the performance
gradually decreasing with $p_T$, reflecting the drop in quark vs gluon
purity shown in Fig.~\ref{fig:feynman}.

\begin{figure}[t]
\centering
\includegraphics[width=0.47\textwidth]{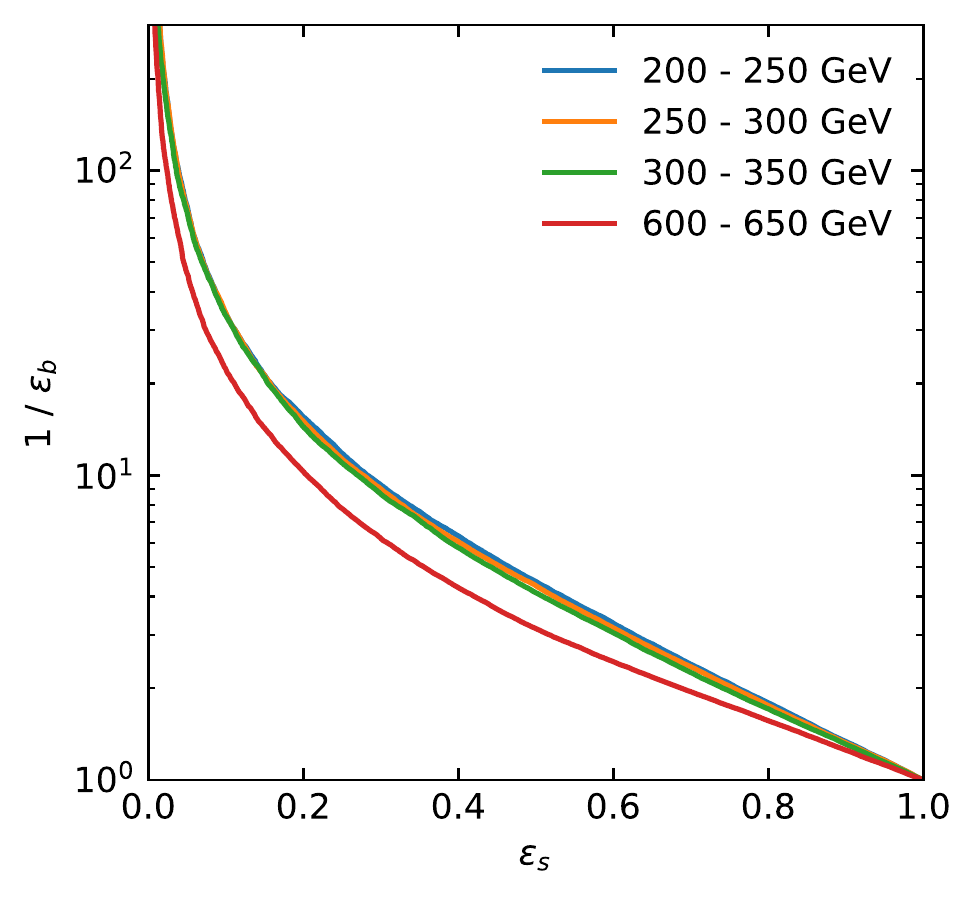}
\caption{ROC curves for the mono-jet samples in non-overlapping jet
  $p_T$ ranges.}
\label{fig:roc_mono}
\end{figure}

\begin{table}[b!]
\begin{footnotesize} 
\begin{minipage}[t]{0.49\textwidth}
\begin{tabular}{c|ccccc}
\toprule
Train & \multicolumn{5}{c}{Test} \\ 
\midrule
& 200-210  &  210-220  &  220-230  &  230-240  &  240-250  \\
\midrule 
200-210  &   0.692   &   0.692   &   0.691    &  0.692   &   0.687    \\   
210-220  &   0.692   &   0.692   &   0.692   &   0.692   &   0.687   \\   
220-230  &   0.692   &   0.692   &   0.692   &   0.692   &   0.688   \\   
230-240  &   0.692   &   0.692   &   0.692   &   0.692   &   0.688   \\   
240-250  &   0.692   &   0.692   &   0.692   &   0.692   &   0.688   \\   
\bottomrule
\end{tabular} 
\end{minipage}
\hspace*{0.06\textwidth}
\begin{minipage}[t]{0.49\textwidth}
\begin{tabular}{c|cccc}
\toprule
Train & \multicolumn{4}{c}{Test} \\ 
\midrule
&  200-250  &  250-300  &  300-350  &  600-650  \\
\midrule
200-250  &   0.691   &   0.683   &   0.674   &   0.604   \\   
250-300  &   0.691   &   0.685   &   0.677   &   0.605   \\   
300-350  &   0.687   &   0.683   &   0.677   &   0.614   \\   
600-650  &   0.630   &   0.638   &   0.646   &   0.631  \\
\bottomrule
\end{tabular} 
\end{minipage}
\end{footnotesize}
\caption{Areas under the ROC curve for the \textsc{LoLa} tagger
  trained and tested on mono-jet samples sliced in $p_{T,j}$. The
  uncertainty on each entry is one to two units on the last shown
  digit.}
\label{tab:mono}
\end{table}

The ROC curves corresponding to the diagonal train and test
combinations of Tab.~\ref{tab:mono}, and their corresponding SI
curves, are shown in Fig.~\ref{fig:roc_mono}. All curves show the same
behavior, with the drop in performance for high-$p_T$ jets visible for
the $600~...~650$~GeV slice. For the actual mono-jet analysis this
implies that quark--gluon discrimination is least efficient when the
analysis focuses on the kinematic regime with the largest missing
energy. However, from Fig.~\ref{fig:kin_mono} we know that for heavy
mediators like a SM-like Higgs this kinematic range is not the most
promising. Instead, we typically analyze the entire $p_{T,j}$
distribution and extract a signal significance from a shape analysis
in the presence of large systematic uncertainties. This is the reason
why we cannot quote a simple significance improvement for the mono-jet
analysis from quark--gluon tagging.  Also for lighter mediators, the
bulk of the $\met$ distribution is what allows us to control the
backgrounds at the required level~\cite{vjets}, and here a systematic
application of quark--gluon tagging may improve our limited
event-level understanding of signal vs background features. On the
other hand, at this level it should be clear that for quark-gluon
discrimination in the presence of detector effects the mono-jet
channel does not provide a useful benchmark.

\section{Di-jet resonances} 
\label{sec:resonance}

As a second application, we study resonances decaying to two
jets. These signal decay jets are usually quark-initiated, while for
relatively light resonances the background will be multi-gluon
production. An interesting aspect of this analysis is that we could,
in principle, use this quark--gluon information already at the trigger
level to enhance the LHC reach in di-jet resonance searches.

We consider an axial vector $Z'$ with a democratic coupling to all
quarks, ignoring the obvious problems with a UV
completion~\cite{uvcomplete}. This resonance might or might not be a
dark matter mediator --- in this study we only consider its decay to
quarks described by the Lagrangian~\cite{Abercrombie:2015wmb}
\begin{align}
\lag_{Z'} = g_{Z'} \sum_q Z'_\mu \, \overline{q} \gamma^\mu\gamma_5 q 
          + \cdots 
\end{align}
The decay to quarks has the benefit that the entire signal only
depends on one kind of coupling, and exactly the coupling we
eventually need to quantitatively analyze mono-jet signals when the
new resonance is a dark matter mediator.  We consider two benchmark
point for the $Z'$ mass, namely $m_{Z'} = 450$~GeV and $m_{Z'} =
750$~GeV, combined with $g_{Z'} = 0.1$, and simulate the signal
and the background with \textsc{Sherpa}2.2.1~\cite{sherpa} to leading
order.  The selection criteria for a standard LHC search are at least
two jets with~\cite{Aaboud:2018fzt}
\begin{align}
p_{T,j_1} > 220 (185)~\gev
\qqquad 
p_{T,j_2} > 85~\gev
\qqquad 
|\eta_j| < 2.8 \; ,
\end{align}
combined with the resonance-inspired requirements
\begin{align}
|y^*| = \frac{|y_{j_1}-y_{j_2}|}{2} < 0.6 (0.3)
\qquad \text{and} \qquad 
\frac{p_{T,j_1} + p_{T,j_2}}{2} = ( 0.6~...~1.4 ) \, p_{T,j_1}   \; .
\end{align}
%

\begin{figure}[t]
 \includegraphics[width=0.49\textwidth]{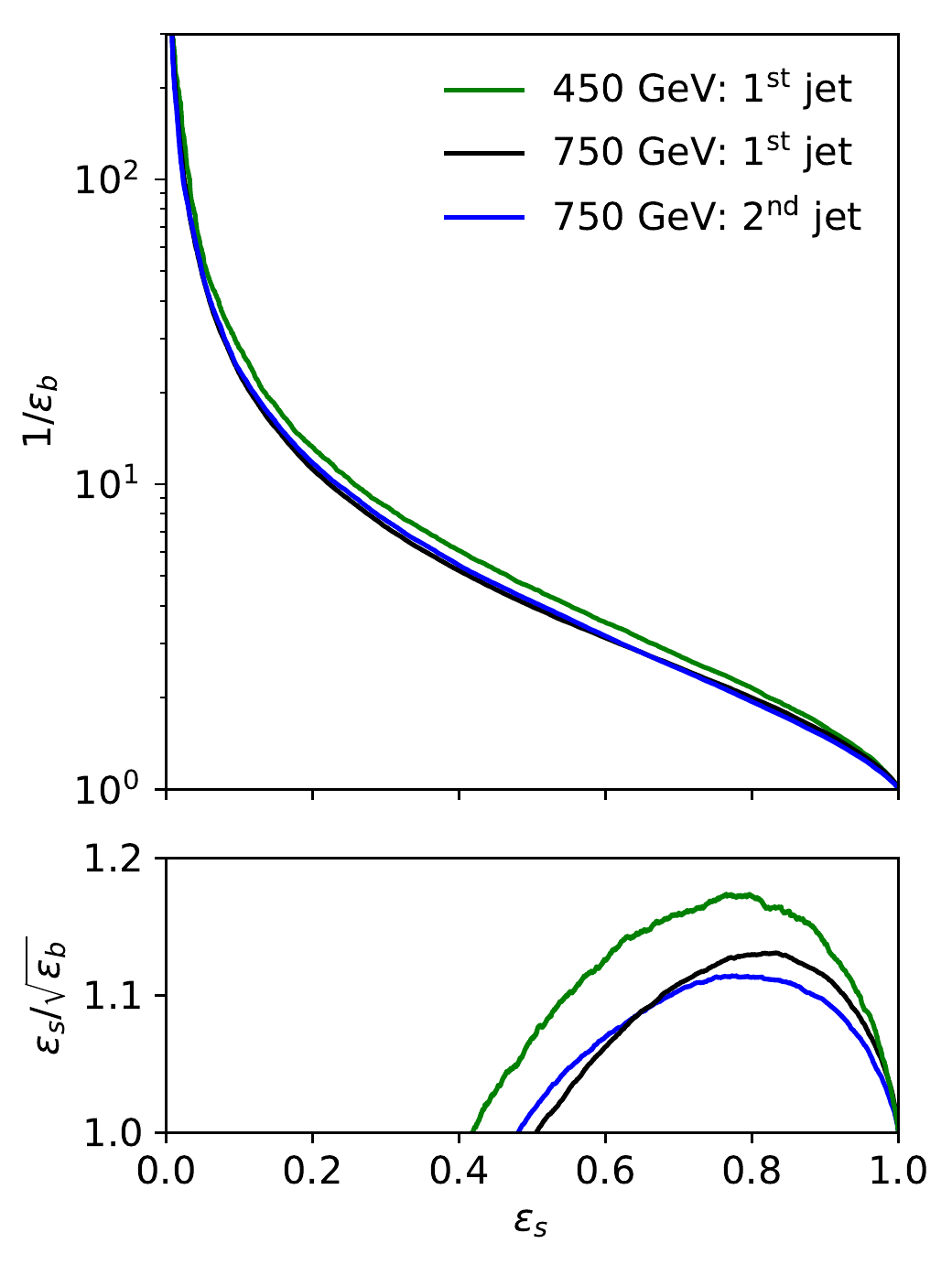}
 \includegraphics[width=0.49\textwidth]{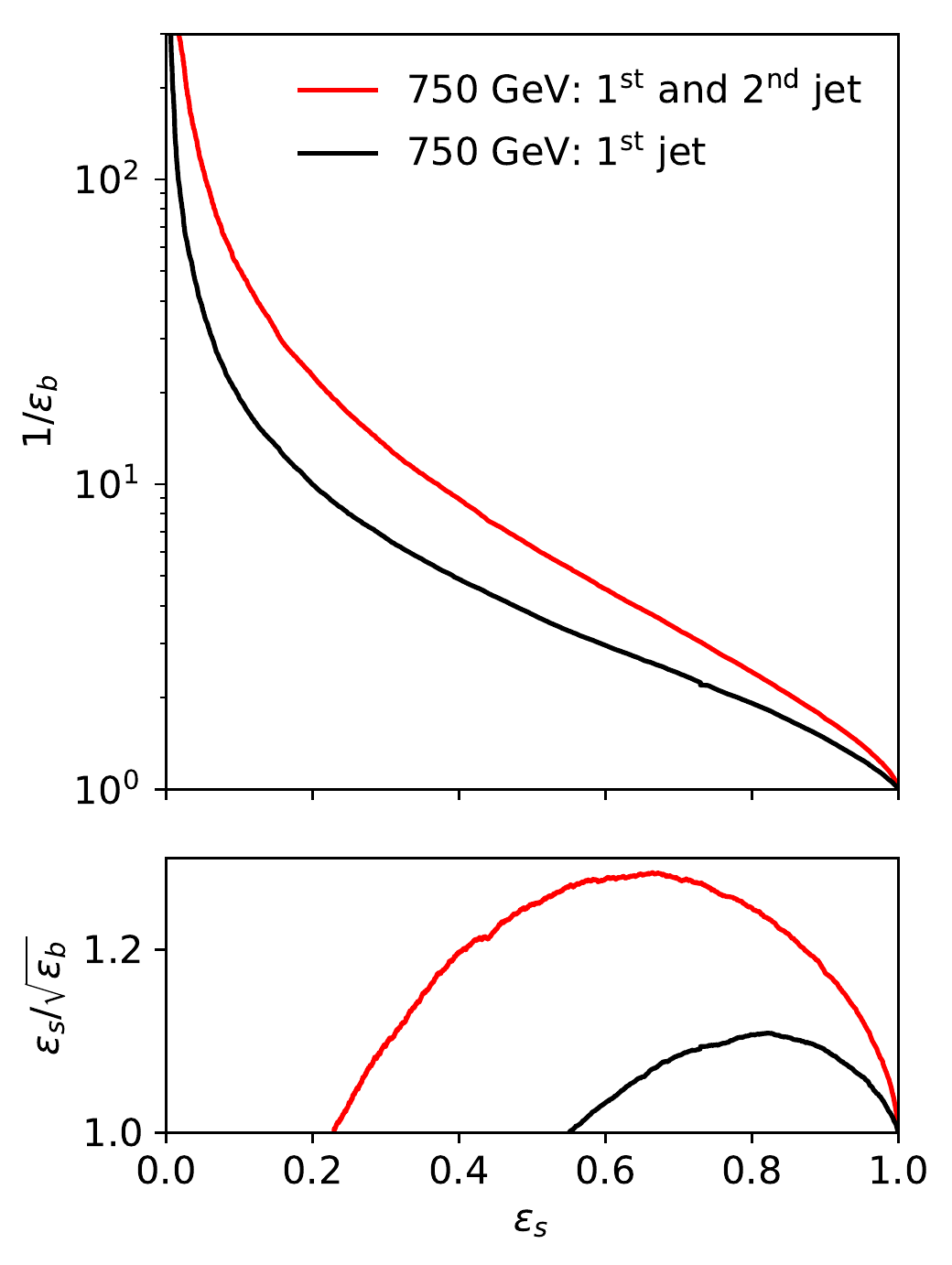}
 \caption{Left: ROC and SI curves for a $250~...~300$~GeV jet from a
   450~GeV $Z'$ resonance and for $300~...~400$~GeV jets from a
   750~GeV $Z'$ resonance, using the model trained on pure
   samples. Right: performance improvement from considering both decay
   jets from a 750 GeV $Z'$ resonance, based on a dedicated training.}
\label{fig:dijets}
\end{figure}

In the left panel of Fig.~\ref{fig:dijets} we first analyze the
leading jet for the low-mass case and both jets for the heavy-mass
case. In both cases we use the pre-trained networks from the pure
samples.  We find that the quark--gluon tagging works slightly better
for lower-mass resonances or lower typical $p_{T,j}$. This has nothing
to do with the signal and is driven by the purity of the QCD
background in this phase space region. The second jet from the light
resonance is comparably soft, which makes it hard to separate it from
QCD radiation without strongly shaping the background.

We also see that, for $m_{Z'} = 750$~GeV the harder jet has more
sensitivity for high signal efficiencies, whereas the second hardest
jet has more sensitivity for lower signal efficiencies. Consequently,
in the right panel of Fig.~\ref{fig:dijets} we show the performance of
a dedicated two-jet \textsc{LoLa} network, combining the network
output from the two jets into an additional set of layers and then
producing the standard di-jet tagging output. As expected, the signal
and the background independently predict two quarks and two gluons, so
the combined network efficiency receives a significant boost. On the
other hand, it is well known that there exist a wealth of observables
which are sensitive to the quark vs gluon nature of jets at the event
level, like additional jet activity. This kind of information is fully
correlated with the quark-gluon tagging of the di-jets, and it is
unlikely that the jet tagging significantly improves the LHC reach
once all event-level observables are considered~\cite{onke}. On the
other hand, these event-level observables are non-trivial to control,
so adding quark-gluon tagging should help controlling the backgrounds.
In that sense, just as for the mono-jet case, our simple significance
estimate is not the whole story. Resonance searches are only partly
limited by statistical significance. Enriching the signal samples with
quarks at an early stage will generally suppress multi-jet
backgrounds. Because trained neural networks are fast, they could be
used already at the trigger level to provide an improved event sample
and to allow for searches in tough phase space regions.

\section{Summary} 
\label{sec:summary}

Quark--gluon separation is one of the hardest problems in contemporary
LHC physics. Technically, is has received a huge boots from machine
learning on low-level observables. Also on the theory side, the
general move towards likelihood-free analyses just comparing fully
simulated and observed events at the detector level circumvents some
of the fundamental QCD problems. In combination, these developments
call for a realistic study of these methods using benchmark signal
processes.

We have extended our \textsc{LoLa} tagger, previously used for top
tagging, to statistically separate quarks from gluons.  For the ideal
case of pure quark and gluon jets we find that detector effects lead
to a degradation of the machine learning results, to a point where a
classic BDT analysis becomes competitive. However, we also remind
ourselves that the standard observables entering the BDT are neither
theoretically nor experimentally preferable and also show non-trivial
correlations. Including charge information in \textsc{LoLa} can be
useful for hard jets. Finally, we have shown that training and testing
the network on sliced of $p_{T,j}$ leads to surprisingly stable
results.

Our first benchmark channel is mono-jet production with a gluon-rich
signal. Subjet information can be added to an otherwise very limited
number of event-level observables. It has the potential to improve
the LHC reach, especially when we use it to understand and control the
entire $p_{T,j}$ distribution. The impact of $p_T$-dependent training
on the systematic uncertainties should be easily controllable.

The second benchmark channel are di-jet resonances with their
quark-rich signal. We find that applying a network trained on pure
samples already improves the reach for relatively light $Z'$ bosons
just using their couplings to quarks. Using our \textsc{LoLa} setup we
find that for hadronically decaying $Z'$ bosons with masses below the
TeV range the quark--gluon discrimination can be useful.

Altogether, we have shown that quark--gluon tagging is a theoretical
and experimental challenge, that deep learning provides competitive
taggers, and that their tagging performance is significantly affected
by detector effects. At the LHC, there exists a range of applications,
both with quark-rich and gluon-rich signals, for which it would be
interesting to see how quark--gluon tagging affects triggering,
background systematics, or the signal extraction in a properly
described experimental setup. Unfortunately, just like weak boson
fusion~\cite{onke} neither mono-jet searches nor di-jet resonance
searches are obvious benchmarks to estimate the impact of quark-gluon
tagging on the LHC reach.

\bigskip
\begin{center} \textbf{Acknowledgments} \end{center}

We are very grateful to Michael Russell for his contributions during
an early phase of this project. We also would like thank Monica
Dunford and Hanno Meyer zu Theenhausen for very useful discussions
about the di-jet channel.  Finally, we acknowledge support by the
state of Baden-W\"urttemberg through bwHPC and the German Research
Foundation (DFG) through grant no INST 39/963-1 FUGG (bwForCluster
NEMO).


\end{document}